\documentclass[12pt,preprint]{aastex}

\usepackage{amssymb,amsmath}

\newcommand{\etal}{et\,al.}
\newcommand{\eg}{e.\,g.}
\newcommand{\ie}{i.\,e.}

\newcommand{\lmax}{l_{\mathrm{max},0}}
\newcommand{\lone}{l_{1,0}}

\newcommand{\plone}{P_{l_{1,0}}}
\newcommand{\qlmax}{Q_{l_{\mathrm{max},0}}}

\newcommand{\neit}{N_{\mathrm{EIT}}}
\newcommand{\rj}{R$_{\mathrm{J}\,}$}
\newcommand{\mr}{m_{R}}

\slugcomment{Accepted for publication by the Astrophysical Journal 05/2003}

\shorttitle{Reflected Light from Giant Planets}
\shortauthors{Jenkins and Doyle}

\begin{document}

\title{Detecting Reflected Light from Close-In Extrasolar Giant
Planets with the \emph{Kepler} Photometer}


\author{Jon M. Jenkins} \affil{SETI Institute/NASA Ames Research
Center, MS 244-30, Moffett Field, CA 94035}
\email{jjenkins@mail.arc.nasa.gov}

\and

\author{Laurance R. Doyle}\affil{SETI Institute, 2035 Landings
Drive, Mountain View, CA 94043} 
\email{ldoyle@seti.org}

\begin{abstract}
NASA's \emph{Kepler Mission} promises to detect transiting Earth-sized
planets in the habitable zones of solar-like stars.  In addition, it
will be poised to detect the reflected light component from close-in
extrasolar giant planets (CEGPs) similar to 51 Peg b.  Here we use the
DIARAD/\emph{SOHO} time series along with models for the reflected
light signatures of CEGPs to evaluate \emph{Kepler's} ability to
detect such planets.  We examine the detectability as a function of
stellar brightness, stellar rotation period, planetary orbital
inclination angle, and planetary orbital period, and then estimate the
total number of CEGPs that \emph{Kepler} will detect over its four
year mission.  The analysis shows that intrinsic stellar variability
of solar-like stars is a major obstacle to detecting the reflected
light from CEGPs.  Monte Carlo trials are used to estimate the
detection threshold required to limit the total number of expected
false alarms to no more than one for a survey of 100,000 stellar light
curves.  \emph{Kepler} will likely detect 100-760 51 Peg b-like
planets by reflected light with orbital periods up to 7 days.

\end{abstract}

\section{Introduction}\label{s:intro}

The discovery of 51 Peg b by \citet{mayorqueloz1995} ignited a
firestorm in the astronomical community.  Eight years later, over 100
extrasolar planets have been found, including multiple planet systems
\citep{butler1999}, and planets in binary systems \citep{cochran1997}. 
The quest for extrasolar giant planets has moved beyond the question
of detecting them to the problem of studying their atmospheres. 
Shortly after the seminal discovery of 51 Peg b, attempts were made to
detect spectroscopically the light reflected from extrasolar planets
in short-period orbits \citep{cameron1999,charb1999}.  To date, no
solid detection of the reflected light component has been reported for
any extrasolar planet, although \citet{charbonneau2002} report the
detection of a drop in the sodium line intensity from the atmosphere
of HD209458b during a transit of its parent star.

The lack of a reflected light component detection is puzzling since a
planet exhibiting a Lambert-like phase function with an albedo similar
to that of Jupiter should be detectable.  The work of
\citet{seageretal2000} provides a possible reason for the lack of
detections: realistic model atmospheres could be significantly less
reflective than would be expected from a Lambert sphere.
Efforts to detect the periodic reflected light components of CEGPs
might be forced to wait for the first generation of space-based
photometers, of which several are scheduled to be launched in the near
future.

The Canadian MOST, the Danish MONS \citep{perryman2000}, and the CNES
mission COROT \citep{schneideretal1998} all promise to study the
miniscule photometric variations indicative of acoustic oscillations
in nearby stars, effectively peering within their hearts to reveal
their internal structure.  These missions will be able to study the
intrinsic stellar variations of the stars they target, much as the
p-modes of the Sun have been studied by ESA's \emph{SOHO Mission}
\citep{frohlichetal1997}.  If the stellar variability does not prove
insurmountable, some of these missions may well be able to detect the
reflected light components of the previously discovered CEGPs.  In
general, the time spent on each target star will not allow these
missions to discover new planets this way, although COROT has the best
chances of doing so, as it surveys a field of stars for several months
at a time, rather than observing a single star at a time.  There are,
however, larger, more ambitious photometric missions on the horizon. 
Both NASA's \emph{Kepler Mission} and ESA's \emph{Eddington Mission}
will be launched in 2007 to search for Earth-sized planets transiting
solar-like stars.  The exquisite photometric precision promised by
these two missions (better than $2\times 10^{-5}$ on timescales of
$\sim$1 day) will not only allow for discovery of transiting planets
and stunning asteroseismology, but might also produce a significant
number of detections of the reflected light from CEGPs.

The reflected light signature of an extrasolar planet appears
uncomplicated at first, much like the progression of the phases of the
moon.  As the planet swings along its orbit towards opposition, more
of its star-lit face is revealed, increasing its brightness.  Once
past opposition, the planet slowly veils her lighted countenance,
decreasing the amount of light reflected toward an observer.  As the
fraction of the visible lighted hemisphere varies, the total flux from
the planet-star system oscillates with a period equal to the planetary
orbital period.  \citet{seageretal2000} showed that the shape of the
reflected light curve is sensitive to the assumed composition and size
of the condensates in the atmosphere of a CEGP. While this presents an
opportunity to learn more about the properties of an atmosphere once
it is discovered, it makes the process of discovery more complex: The
reflected light signatures are not as readily characterized as those
of planetary transits, so that an ideal matched filter approach does
not appear viable.  The signatures from CEGPs are small ($<$100 ppm)
compared to the illumination from their stars, requiring many cycles
of observation to permit their discovery.  This process is complicated
by the presence of stellar variability which imposes its own
variations on the mean flux from the star.  Older, slowly rotating
stars represent the best targets.  They are not as active as their
younger counterparts, which are prone to outbursts and rapid changes
in flux as star spots appear, evolve and cross their faces.  In spite
of these difficulties, a periodogram-based approach permits the
characterization of the detectability of CEGPs from their reflected
light component.


Our study of this problem began in 1996 in support of the proposed
\emph{Kepler Mission}\footnote{\url{www.kepler.arc.nasa.gov}} to the
NASA Discovery Program \citep{borucki1996,doyle1996}.  That study used
measurements of solar irradiance by the Active Cavity Radiometer for
Irradiance Monitoring (ACRIM) radiometer aboard the \emph{Solar
Maximum Mission} (\emph{SMM}) \citep{willson1991}, along with a model
for the reflected light signature based on a Lambert sphere and the
albedo of Jupiter.  Here we significantly extend and update the
previous preliminary study using measurements by the Dual Irradiance
Absolute Radiometer (DIARAD), an active cavity radiometer aboard the
Solar Heliospheric Observatory (\emph{SOHO}) \citep{frohlichetal1997}
along with models of light curves for 51 Peg b\textendash like planets
developed by \citet{seageretal2000}.  For completeness, we include
Lambert sphere models of two significantly different geometric
albedos, $p$=0.15 and $p$=2/3.  The \emph{SOHO} data are relatively
complete, extend over a period of 5.2 years, are evenly sampled at 3
minutes, a rate comparable to that for \emph{Kepler's} photometry (15
minutes), and have the lowest instrumental noise of any comparable
measurement of solar irradiance.  \citet{seageretal2000} provide an
excellent paper describing reflected light curves of CEGPs in the
visible portion of the spectrum.  However, they do not consider the
problem of detecting CEGP signatures in realistic noise appropriate to
high precision, space-based photometers.

The current article complements our study of the impact of solar-like
variability on the detectability of transiting terrestrial planets for
\emph{Kepler} \citep{jenkins2002}.  The observational noise
encountered in the detection of transiting small planets and in the
detection of the reflected light component of CEGPs is the same: only
the shape, timescale and amplitude of the signal of interest have
changed.  We have also conducted a more thorough analysis of the false
alarm rates and the requisite detection thresholds than that performed
in our preliminary study and have developed a detection scheme to
accommodate the non-sinusoidal nature of the reflected light curves
produced by CEGPs.  In this paper, our focus is on the \emph{Kepler
Mission} due to our familiarity with its design parameters and its
expected instrumental noise component, although the results should
apply to other missions with similar apertures, instrumental noise,
and mission durations.

In this study, we analyze different combinations of model planetary
atmospheres, stellar rotation periods, and stellar apparent
magnitudes, examining the detectability of each case over a range of
orbital inclinations, $I$, from edge\textendash on ($I=90^{\circ}$) to
near broadside ($I=10^{\circ}$), and over orbital periods, $T_p$, from 2
to 7 days.  The brightnesses considered for the stars range from
$\mr$=9.0 to $\mr$=15.0, which dictate the corresponding shot and
instrument noise for the \emph{Kepler} target stars.  To simulate the
reflected component of the light curve we use two atmospheric models
developed by \citet{seageretal2000}: one for clouds with a mean
particle radius, $\bar{r}$, of 0.1 $\mu$m consisting of a mixture of
Fe, MgSiO$_3$, and Al$_{2}$O$_3$, and the same mixture, but for clouds
with $\bar{r}=1.0$ $\mu$m.  We also consider two Lambert sphere
models, one with maximum reflectivity, and one with a geometric albedo
of 0.15 (corresponding to the case of Mars).  Different models of
stellar variability are considered, all based on the
DIARAD/\emph{SOHO} time series, which were resampled and scaled to
obtain synthetic light curves for stars with rotation periods between
5 and 40 days as per Jenkins (2002).  For each set of parameters, a
periodogram analysis yields the expected detection statistic for a 1.2
Jupiter radius (R$_{\mathrm{J}}$) planet.  The appropriate detection
thresholds and resulting detection rates are determined from Monte
Carlo runs using the same detection procedure applied to White
Gaussian Noise (WGN) sequences.  The resulting detection rates are
averaged over all orbital inclinations and over the expected
distribution of CEGP orbital periods, and are then used in conjunction
with a model distribution of main sequence stars in \emph{Kepler's}
field of view (FOV) to estimate the total number of CEGPs detected by
reflected light.  The results indicate that \emph{Kepler} should
detect 100-760 CEGPs in orbits up to 7 days in period around old,
quie, solar-like stars.  These detections will not occur in the first
weeks of the mission due to the low amplitudes of the planetary
signatures.  Rather, they will accumulate steadily over the course of
the mission.

The paper is organized as follows: We present the DIARAD/\emph{SOHO}
measurements of solar variability in \S\ref{s:diarad}, followed by a
discussion of \citet{seageretal2000}'s light curve models for the
reflected light component from CEGPs in \S\ref{s:atmosmodels}.  A
summary of the \emph{Kepler Mission} is given in \S\ref{s:kepler}. 
Section \S\ref{s:besancon} describes the galactic model for the
distribution of \emph{Kepler's} target stars used to optimize the
proposed detection algorithm and analyze its performance.  Our
approach to detecting strictly periodic signals in noise and setting
detection thresholds and assessing detection rates is given in
\S\ref{s:detapproach}.  Monte Carlo experiments conducted to establish
false alarm rates and the requisite detection thresholds are discussed
in \S\ref{s:montecarlo}.  The expected number of detections is
presented in \S\ref{s:numdetections}.  A discussion of sources of
confusion and methods to reject false positives is given in
\S\ref{s:confusionanddiscrimination}.  We conclude in
\S\ref{s:conclusions} by summarizing the findings and giving
suggestions for future work.

\section{The DIARAD/\emph{SOHO} Observations}\label{s:diarad}

In order to study the capabilities of missions such as \emph{Kepler},
we take measurements from the DIARAD instrument aboard the \emph{SOHO}
spacecraft as a proxy for all solar-like stars.  DIARAD is a
redundant, active-cavity radiometer aboard \emph{SOHO} that measures
the white-light irradiance from the Sun every 3 minutes
\citep{frohlichetal1997}.  The DIARAD measurements considered here
consist of 5.2 yr of data that begin near solar minimum in January,
1996 and extend to March, 2001, just past solar maximum.  The data are
not pristine: there are gaps in the data set, the largest of which
lasts 104 days, and there are obvious outliers in the data. 
Nevertheless, the DIARAD time series is the most uniformly-sampled,
lowest noise data set available.  Once it is binned to \emph{Kepler's}
sampling rate (4 hr$^{-1}$), fully 83\% of the data samples are
available (62$\%$ of the missing points are represented by the three
largest data gaps).  We've taken the liberty of removing the obvious
outliers and have filled in the missing data as per
\citet{jenkins2002} in such a way as to preserve the correlation
structure of the underlying process.

Ground-based observations show that solar-type stars rotating faster
than the Sun are more magnetically active, increasing the photometric
variability over a range of timescales.  These observations generally
consist of sparse, irregularly sampled time series with usually no
more than one measurement per star per night.  Thus, it is difficult
to use these observations to study the distribution of variability on
timescales shorter than a few days.  They do, however, provide an
indication of the appropriate scaling relation to use on timescales $>
1$ day.  Figure 7 of \citet{radicketal1998} indicates that photometric
variability, $\sigma_{phot}$, on time scales shorter than a year is
related to the chromospheric activity level parameter,
$\rm{R}'_{\rm{HK}}$, by a power law with exponent 1.5.  Other
observations \citep{noyesetal1984} suggest that $\rm{R}'_{\rm{HK}}$ is
approximately inversely proportional to stellar rotation period,
$P_{rot}$, so that
\begin{equation}
	\sigma_{phot}\propto P_{rot}^{-1.5}.
\label{eq:scalefcn}
\end{equation}
This scaling relation is used to scale the variability of the DIARAD
measurements on timescales longer than 2 days.  

The DIARAD measurements themselves represent a means by which the
timescale-dependent response of solar-like stars to increased magnetic
activity can be estimated.  At solar maximum (with high magnetic
activity levels), variability at long timescales increases
significantly relative to solar minimum, while it remains
comparatively constant at timescales of hours [see Fig.  2 of
\citet{jenkins2002}].  Synthetic time series can be generated by
transforming the DIARAD time series into the wavelet domain, scaling
each timescale component by a factor which is one at the shortest
timescales and that ramps up to the value indicated by the
ground-based measurements for timescales $\ge 2.66$ days, followed by
resampling the time series onto an appropriate grid
\citep{jenkins2002}.  This procedure represents our best estimate of
how the stellar rotation period should affect the photometric
variability of solar-like stars.  We do not expect this model to be
accurate over a wide range of stellar types.  It probably is only
indicative of the expected effects over stellar types near the Sun
(G1$-$G4).  Warmer, late-type stars generally exhibit less spotting and
consequently, lower $\sigma_{phot}$, while cooler, late-type stars
exhibit more spotting and higher $\sigma_{phot}$ for a given $P_{rot}$
\citep[see, \eg,][]{messinaetal2001}.  Warmer, late-type stars,
however, are also larger, requiring a larger planet to achieve the
same S/N for a given photometric variability, while cooler, late-type
stars are smaller, mitigating the increased variability for a given
size planet to some degree.  This analysis does not include the
effects of flare events, which exhibit transient signatures on
timescales of minutes (more frequently) to a few hours (more rarely),
the frequency of which increases significantly for rapid rotators.

\begin{figure}
\plottwo{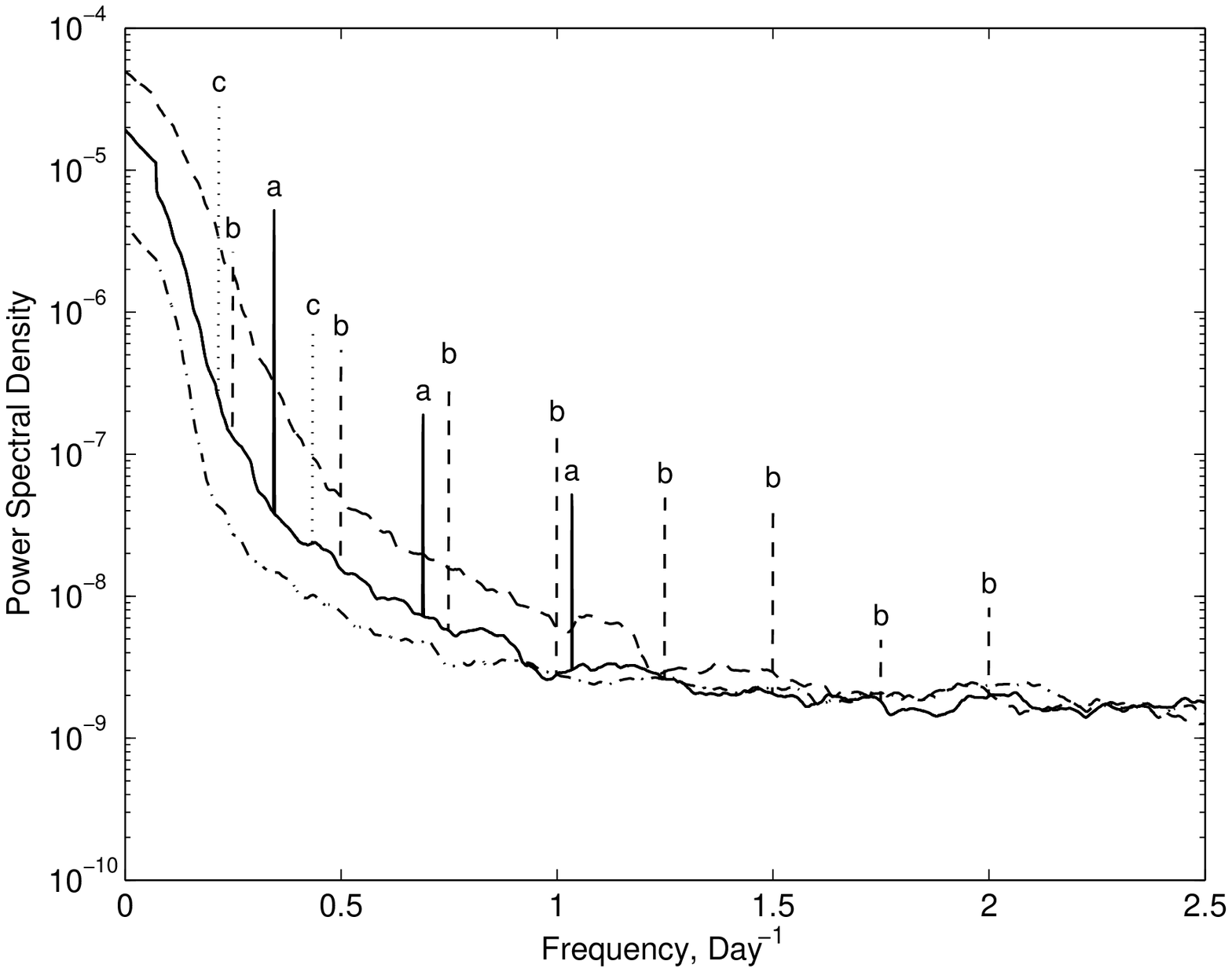}{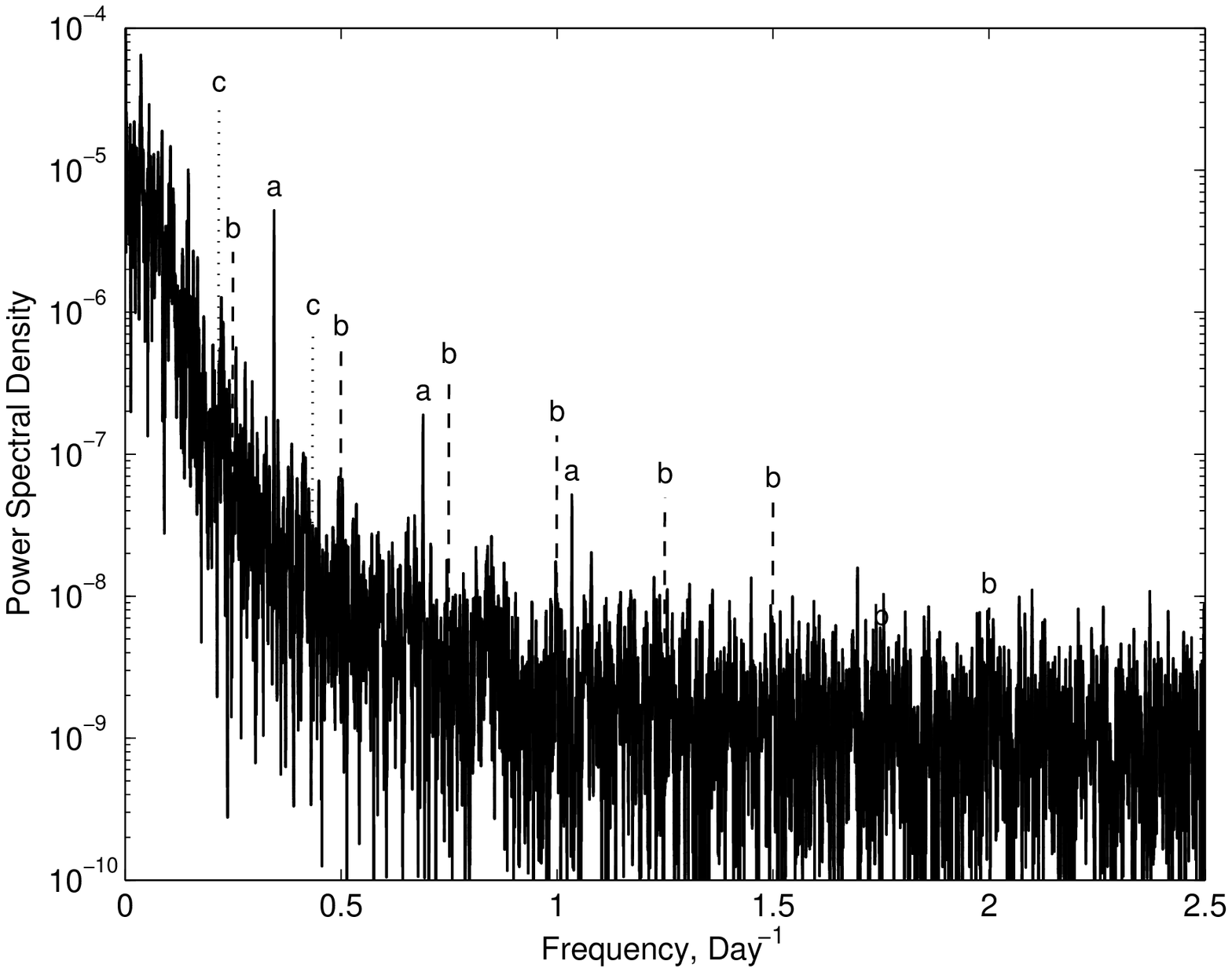} \caption[F1.eps]{\small{Power spectral
density (PSD) estimates for solar-like variability and signatures of
three extrasolar giant planets.  Panel a) displays Hanning-windowed
periodograms for a combination of the first 4 yr of the DIARAD data
set and three reflected light CEGP signatures.  The three planetary
signatures are for 1.2 R$_{\mathbf{J}}$ planets with atmospheres
composed of 1.0 $\mu$m particles in a 4 day orbit, a planet with 0.1
$\mu$m particles in a 2.9 day orbit, and a 4.6 day, albedo $p=2/3$,
Lambert sphere.  The planetary signatures consist of impulse trains
with their harmonic components denoted by `a's, `b's and `c's,
respectively.  The noise fluctuations in PSD estimates are quite
evident in panel a).  Three solar-like PSDs are displayed in panel b),
along with a combination of these same planetary signatures and a 26.6
day period, solar-like star.  The stellar PSDs have been smoothed by a
21-point moving median filter (0.015 Day$^{-1}$ wide) followed by a
195-point moving average filter (0.14 Day$^{-1}$ wide) to illustrate
the average background noise.  This is the procedure used by the
proposed detector to estimate the background stellar PSDs prior to
whitening the observed periodograms.  The solid curve corresponds to
the DIARAD data in panel a ($P_{rot}=26.6$ days), while the dashed and
dash-dotted curves are for solar-like stars with rotation periods of
20 and 35 days, respectively, demonstrating the dependence of stellar
variability on stellar rotation period.  Three harmonic components of
the planet with 0.1 $\mu$m particles (solid lines topped with `a's)
are visible above the noise in panel a), while seven components of the
planet with 1.0 $\mu$m particles are visible (dashed lines topped with
`b's).  Only two components (dotted lines topped with `c's) of the
$p=2/3$ Lambert sphere are visible.  Thus, it should be possible to
constrain the particle size distribution and composition of a CEGP
atmosphere by the number of detected Fourier components.  On this
scale, the planetary signatures appear as vertical line segments,
though they are actually distributed over a few frequency bins.  }}
\label{fig:Pfig}
\end{figure}

\placefigure{fig:Pfig}

Figure \ref{fig:Pfig}a shows a portion of the power spectral density,
PSD, for the Sun from a frequency of 0 to 2.5 day$^{-1}$.  Figure
\ref{fig:Pfig}b shows a smoothed version of the same solar PSD along
with PSDs for solar-like stars with rotation periods of 20 and 35
days.  The stellar PSDs in Figure \ref{fig:Pfig}b have been smoothed
by a 21-point moving median filter (0.015 Day$^{-1}$ wide) followed by
a 195-point moving average filter (0.14 Day$^{-1}$ wide) to emphasize
the average background noise.  The effect of decreasing $P_{rot}$ is
to increase the low frequency noise and the frequency at which the PSD
rolls off.  The PSD for the Sun falls rapidly from 0 day$^{-1}$ to
0.25 day$^{-1}$, then gradually flattens out so that it is nearly
level by 1 day$^{-1}$.  Most power occurs at frequencies less than 0.1
days$^{-1}$, corresponding to the rotation of sunspots and solar-cycle
variations \citep{frohlich1987}.  On time scales of a few hours to a
day, power is thought to be dominated by convection-induced processes
such as granulation and supergranulation \citep{rabellosoares1997,
andersenetal1998}.  At $\sim$288 day$^{-1}$, beyond the axis limits of
the figure, the so-called p-modes corresponding to acoustic resonances
can be observed with typical amplitudes of 10 ppm .


\section{Atmospheric Models and Synthetic Reflected Light
Signatures}\label{s:atmosmodels} 

Motivated by the upcoming microsatellite missions for studying
asteroseismology, \citet{seageretal2000} investigated the optical
photometric reflected light curves expected for CEGPs.  Their model
code solves for the emergent planetary flux and temperature-pressure
structure in a self-consistent fashion.  The solution is found while
simultaneously satisfying hydrostatic equilibrium, radiative and
convective equilibrium, and chemical equilibrium in a plane-parallel
atmosphere, with the impinging stellar radiation setting the upper
boundary conditions.  A three-dimensional Monte Carlo code computes
the photometric light curves using the solution for the atmospheric
profiles.  For the Gibbs free energy calculations, they include 27
elements, 90 gaseous species and four solid species.  These include
the most important species for brown dwarfs and cool stars.  The
condensates, solid Fe, MgSiO$_{3}$, and Al$_{2}$O$_{3}$ are likely to
be present in the outer atmospheres of the CEGPs.  Four mean sizes of
condensate are considered, spanning a large fraction of the range of
sizes observed in planetary atmospheres in the solar system: 0.01-10
$\mu$m.  \citet{seageretal2000} emphasize that their study is a
preliminary one, as significant improvements can be made in cloud
modeling, atmospheric circulation and heat transport, photochemistry,
and the inclusion of other condensates.  Indeed, work to incorporate
more realistic physics into these models is ongoing
\citep{greenetal2002}\, but has not resulted in significant revisions
in the general shape or amplitudes of the reflected light signatures
(Sara Seager 2002, personal communication).  Therefore, the published
photometric light curves represent a sufficient starting point for
investigating the detectability of the reflected light signatures of
CEGPs and are significantly better than what can be obtained by using
a Jupiter-like albedo and a simple analytic model such as a Lambert
sphere.

\citet{seageretal2000} find that the amplitude of the reflected light
curves from CEGPs is significantly lower than that due to a Lambert
sphere, which yields a signal as high as 83 ppm for a 1.2
R$_{\mathrm{J}}$ planet in a 0.051 AU orbit about a G2 star.  Instead,
they predict a peak flux of 22 ppm for an atmosphere consisting of a
distribution of particles with a mean radius of 0.1 $\mu$m in a
uniform cloud consisting of a mixture of Fe, MgSiO$_3$, and
Al$_{2}$O$_3$, and at a wavelength of 0.55 $\mu$m.  The scattering
from these particles is at the upper limit of the Rayleigh regime, so
the resulting light curve is relatively smooth.  For an atmosphere
composed of $\bar{r}$=1.0 $\mu$m particles, the scattering is well
into the Mie regime, resulting in a strong central peak of 52 ppm
centered at opposition, mainly due to backscattering from the
MgSIO$_{3}$ particles at low phase angles and forward diffraction of
all particles at higher phase angles, creating the steep wings at
intermediate phase angles.  \citep{seageretal2000} also remark that an
atmosphere with stratified cloud layers would likely result in
significantly higher flux reflected from the planet, as the top-level
cloud would consist of MgSiO$_{3}$ because of its cooler condensation
curve.  For example, for $\bar{r}$=0.01 $\mu$m particles consisting of
a uniform mixture of condensates, the amplitude of the reflected light
signature is very low, ~0.2 ppm, while for pure MgSIO$_{3}$ it is ~100
times stronger.  The case of $\bar{r}$=10.0 $\mu$m results in higher
amplitudes for the reflected light signatures, both for the mix of the
four condensates, and for pure MgSiO$_{3}$.  \citet{seageretal2000}
consider particle sizes found in planetary atmospheres in the solar
system.  A $\bar{r}=$0.01 $\mu$m particle size corresponds to the haze
layer above the main cloud layer in the atmosphere of Venus, which in
contrast to the haze, consists of particles of size ~1 $\mu$m
\citep{knollenbergetal1980}, while the cloud particles in Jupiter's
upper atmosphere span 0.5-50 $\mu$m \citep{westetal1986}.  We
therefore take the light curves for the uniform mixture with
$\bar{r}$=0.1 $\mu$m and $\bar{r}$=1.0 $\mu$m as conservative cases
for the purpose of examining photometric detectability.

The light curves given by \citet{seageretal2000} can be scaled to
planet-star separations different from that of 51 Peg by noting that
the reflected light component amplitude is inversely proportional to
the square of the star-planet separation.  The authors caution that
the planetary atmosphere and its cloud structure are sensitive to the
insolation experienced by the planet so that this scaling law may not
produce accurate results much beyond 0.05 AU or inside of 0.04 AU.
They suggest, however, that the scaling law produces rough estimates
at planet-star separations as high as 0.12 AU. To be complete, we also
consider the detectability of CEGPs whose reflected light components
are better modeled as Lambert spheres, with geometric albedos $p=0.15$
(roughly corresponding to that of Mars) and $p=2/3$ (the maximum). 
These light curves would likely arise from a cloudless atmosphere with
a uniformly-distributed absorbing gas controlling the albedo.

\section{The Kepler Mission}\label{s:kepler}

\emph{Kepler}, a recently selected NASA Discovery Mission, is designed
to detect Earth-sized planets orbiting solar-like stars in the
circumstellar habitable zone.  More than $100,000$ target stars will
be observed in the constellation Cygnus continuously for at least 4 yr
at a sampling rate of 4~hr$^{-1}$ \citep{borucki1997}. 
\emph{Kepler's} aperture is 0.95~m allowing $2.21\times 10^{8}~e^{-}$
to be collected every 15 minutes for a G2, $\mr=12$ dwarf star with a
shot noise of 67 ppm.  The instrument noise itself should be $\sim$~31
ppm over this same duration.  This value is based on extensive
laboratory tests, numerical studies and modeling of the \emph{Kepler}
spacecraft and photometer
\citep{kochetal2000,jenkinsetal2000,remundetal2001}.  The values in
Table 3 of \citet{kochetal2000} support this level of instrumental
noise from a high-fidelity hardware simulation of \emph{Kepler's}
environment, while the numerical studies of \citet{remundetal2001} are
based on a detailed instrumental model.  This model includes noise
terms such as dark current, read noise, amplifier and electronics
noise sources, quantization noise, spacecraft jitter noise, noise from
the shutterless readout, cosmic ray hits, radiation damage accumulated
over the lifetime of the mission, and the effects of charge transfer
efficiency.

To simulate the combined effects of the shot noise and instrumental
noise for $Kepler$, white Gaussian noise (WGN) sequences were added to
the DIARAD time series with a standard deviation equal to the square
root of the combined shot and instrumental variance for a star at a
given magnitude less the square of the DIARAD instrumental uncertainty
[$0.1~W~m^{-2}$ in each 3 minute DIARAD measurement (Steven Dewitte
1999, personal communication)].  For example, the combined
instrumental and shot noise for a $\mr$=12 star at the 15 minute level
is $\sim$74 ppm, but the DIARAD instrumental noise is $\sim$33 ppm, so
the appropriate standard deviation for the WGN sequence is 66 ppm.

The \emph{Kepler Mission} should not suffer from large time gaps. 
Roll maneuvers are planned about every 90 days to reorient the
sunshade and the solar panels, resulting in a loss of $\sim$1\% of the
total data.  While the simulations discussed in \S\ref{s:detapproach}
do not include the effect of the missing data, it should be small and
can be accommodated directly into a tapered spectrum estimate as per
\citet{waldenetal1998}.

\section{A Galactic Model for the Distribution of \emph{Kepler} 
Target Stars}
\label{s:besancon}

Along with characterizations of stellar variability as a function of
stellar rotation period, $P_{rot}$, and a characterization of the
observation noise for \emph{Kepler}, we require a model for the
distribution of \emph{Kepler's} main-sequence target stars as
functions of apparent magnitude, spectral type, and age.  Combined
with a characterization of the detectability of CEGPs with respect to
apparent brightness and $P_{rot}$, the stellar distribution allows for
the performance of the proposed detector to be optimized and evaluated
(\S\ref{s:detapproach} and \S\ref{s:montecarlo}).  In addition, a
model of the distribution of dim background stars in the FOV permits
an analysis of the problem of confusion
(\S\ref{s:confusionanddiscrimination}).

Following \citet{batalha2002}, we make use of galactic models made
publicly available by the Observatoire de
Besan\c{c}on\footnote{\url{http://www.obs.-besancon.fr/modele/modele.ang.html}}
\citep[see, \eg,][]{robincreze1986,haywoodetal1997a,haywoodetal1997b}
to obtain expected main sequence starcounts as a function of apparent
magnitude, spectral type and age.  The USNO-A2.0 database yields
223,000 stars to $\mr$=14.0 in the 106 square degrees of
\emph{Kepler's} FOV (David Koch 2001, personal communication).  This
establishes an appropriate mean extinction of $\sim$1.0 mag
$\mathrm{kpc}^{-1}$ for the Besan\c{c}on model.  We note, however,
that the bandpass for \emph{Kepler} extends from $\sim$0.45 to
$\sim$0.85 $\mu$m, which is far wider than the bandpasses available
for the Besan\c{c}on models.  For the purpose of counting stars, using
the R band should reflect the number of stars of greatest interest,
but may tend to undercount the number of late main sequence stars. 
The age-rotation relation formulated by \citet{kawaler1989} then
permits us to rebin the starcounts obtained from the Besan\c{c}on
model with respect to rotation period rather than stellar age, for
which we can evaluate the expected detection rates.  This relation is
given by
\begin{equation}
	\log(P_{rot})=0.5\log(t_{0})+0.390(B-V)+0.824,
\label{eq:agerotation}
\end{equation}
where $P_{rot}$ is stellar rotation period in days, $t_{0}$ is stellar
age in Gyr, and B and V are the blue and visible photometric
brightnesses in the Johnson UBVRI system, respectively.  For this
exercise, the apparent magnitudes were binned into 1 magnitude
intervals with central values from $\mr$=9.5 to 14.5, the spectral
type bins were centered on spectral types B5, A5, F5, G5, K5, and M5,
and the rotation periods were binned into 5 day intervals from 5 to 40
days.  Stars rotating with periods outside this range were set to the
respective edge bin values.

\placetable{nstartbl}
\placetable{nCEGPtbl}
\placetable{ntranCEGPtbl}

Table \ref{nstartbl} gives the number of stars in each spectral type
and apparent magnitude bin.  The Observatoire de Besan\c{c}on galactic
model estimates that there are $\sim$80,000 main sequence stars to
$\mr$=14.0 and $\sim$220,000 main sequence stars to $\mr$=15.0 in
\emph{Kepler's} FOV. Other models exist that predict higher fractions
of main-sequence stars (David Koch 2002, personal communication), so
that this is a reasonably conservative starting point.  There are 14
extrasolar planets currently known with orbital periods less than 7
days: four with periods of nearly 3.0 days, four with periods of
$\sim$3.5 days, and the remaining six are approximately uniformly
distributed between $P$=4 and 6.4 days.  About 0.75\% of solar-like
stars possess planets with periods between 3 and 5 days
\citep{butler1999}, which we scale to 0.875\% since two of the CEGPs
for our model distribution have periods greater than 5 days.  Taking
this value for the fraction of target stars that possess CEGPs, we
obtain the results listed in Table \ref{nCEGPtbl}, for a total of 693
planets to $\mr$=14.0 and 1,807 planets to $\mr$=15.0.  Of these,
$\sim$10\% should exhibit transits, as given in Table
\ref{ntranCEGPtbl}.  The photometric signals from these transiting
planets will be huge compared to the measurement noise, so that
virtually all of these planets whose parent stars are observed by
\emph{Kepler} will be detected.  Thus, there should be $\sim$44 CEGPs
to $\mr$=14.0 and 181 CEGPs to $\mr$=15.0 discovered within the first
several weeks of observation.  The question addressed throughout the
remainder of this paper is, how many additional planets should
\emph{Kepler} be able to detect by reflected light?

\section{Detection Approach}\label{s:detapproach}
The detection of reflected light signatures of non-idealized model
atmospheres such as those predicted by \citet{seageretal2000} is more
complicated than for the signature of a Lambert sphere.  The power
spectrum of any periodic waveform consists of a sequence of evenly
spaced impulses separated by the inverse of the fundamental period. 
For a Lambert sphere, over 96\% of the power in the reflected light
component is contained in the fundamental (aside from the average flux
or DC component, which is undetectable against the stellar background
for non-transiting CEGPs).  Thus, detecting the reflected light
signature of a Lambert sphere can be achieved by forming the
periodogram of the data, removing any broadband background noise, and
looking for anomalously high peaks.  In contrast, the power of the
Fourier expansions of Seager \etal's model CEGP light curves at high
orbital inclinations is distributed over many harmonics in addition to
the fundamental due to their non-sinusoidal shapes (see Fig. 
\ref{fig:Pfig}).  How does one best search for such a
signal?\footnote{A key point in searching for arbitrary periodic
signals, or even pure sinusoids of unknown frequency is that no
optimal detector exists \citep{kay1998}.  The most prevalent approach
is to use a generalized likelihood ratio test which forms a statistic
based on the maximum likelihood estimate of the parameters of the
signal in the data.  Such a detector has no pretenses of optimality,
but has other positive attributes and often works well in practice.}


As in the case of a pure sinusoid, a Fourier-based approach seems most
appropriate, since the Fourier transform of a periodic signal is
strongly related to its Fourier series, which parsimoniously and
uniquely determines the waveform.  Unlike the case for ground-based
data sets that are irregularly sampled and contain large gaps,
photometric time series obtained from space-based photometers like
\emph{Kepler} in heliocentric orbits will be evenly sampled and nearly
complete.  This removes much of the ambiguity encountered in power
spectral analysis of astronomical data sets collected with highly
irregular or sparse sampling.  Thus, power spectral analyses using
Fast Fourier Transforms (FFTs) simplify the design of a detector.  For
the sake of this discussion, let $x(n)$ represent the light curve,
where $n\in\{0,\ldots,N-1\}$ is an $N$-point time series with a
corresponding discrete Fourier transform (DFT) $X(k)$,
$\omega=2\pi\,k/N$ is angular frequency, and $k\in\{0,\ldots,N-1\}$). 
The phase of the light curve is a nuisance parameter from the
viewpoint of detecting the planetary signature and can be removed by
taking the squared magnitude of the DFT, $P_{X}(k)=|X(k)|^{2}$, which
is called the periodogram of the time series $x(n)$.  In the absence
of noise, if the length of the observations were a multiple of the
orbital period, $T_{p}$, then the periodogram would be zero everywhere
except in frequency bins with central frequencies corresponding to the
inverse of the orbital period, $f_{0}=T_{p}^{-1}$, and its multiples. 
If the length of the observations is not an integral multiple of the
orbital period, the power in each harmonic is distributed among a few
bins surrounding the true harmonic frequencies, since the FFT treats
each data string as a periodic sequence, and the length of the data is
not consonant with the true orbital period.  The presence of wide-band
measurement noise assures that each point in the periodogram will have
non-zero power.  Assuming that the expected relative power levels at
the fundamental and the harmonics are unknown, one can construct a
detection statistic by adding the periodogram values together that
occur at the frequencies expected for the trial period $T_{p}$, and
then threshold the summed power for each trial period so that the
summed measurement noise is not likely to exceed the chosen threshold. 
The statistic must be modified to ensure that it is consistent since
longer periods contain more harmonics than shorter ones, and
consequently, the statistical distribution of the test statistics
depends on the number of assumed harmonics.  This is equivalent to
fitting a weighted sum of harmonically related sinusoids directly to
the data.  \citet{kay1998} describes just such a generalized
likelihood ratio test (GLRT) for detecting arbitrary periodic signals
in WGN assuming a generalized Rayleigh fading model.\footnote{In the
Rayleigh fading model for a communications channel, a transmitted
sinusoid experiences multipath propagation so that the received
signal's amplitude and phase are distorted randomly.  A sinusoid of
fixed frequency can be represented as the weighted sum of a cosine and
a sine of the same frequency, with the relative amplitudes of each
component determining the phase.  If both component amplitudes have a
zero mean, Gaussian distribution, then the phase is uniformly
distributed and the amplitude of the received signal has a Rayleigh
distribution.  The generalized Rayleigh fading model consists of a set
of such signals with harmonically related frequencies to model
arbitrary periodic signals.}

The approach we consider is similar; however, we assume the signals
consist of no more than seven Fourier components, and we relax the
requirement that the measurement noise be WGN. This is motivated by
the observation that the model light curves developed by
\citet{seageretal2000} are not completely arbitrary and by the fact
that the power spectrum of solar-like variability is very red: most of
the power is concentrated at low frequencies.  At low inclinations,
the reflected light curves are relatively smooth and quasi-sinusoidal,
exhibiting few harmonics in the frequency domain.  At high
inclinations, especially for the $\bar{r}$=1.0 $\mu$m model, the
presence of a narrow peak at opposition requires the presence of about
seven harmonics in addition to the fundamental (above the background
solar-like noise).  Another GLRT approach would be to construct
matched filters based directly on the atmospheric models themselves,
varying the trial orbital period, inclination, mean particle size,
etc.  A whitening filter would be designed and each synthetic light
curve would be ``whitened'' and then correlated with the ``whitened''
data.\footnote{For Gaussian observation noise and a deterministic
signal of interest, the optimal detector consists of a whitening
filter followed by a simple matched filter detector \citep{kay1998}. 
The function of the whitening filter is to flatten the power spectrum
of the observation noise so that filtered data can be characterized as
white Gaussian noise.  Analysis of the performance of the resulting
detector is straightforward.  For the case of non-Gaussian noise, the
detector may not be optimal, but it is generally the optimal linear
detector, assuming the distribution of the observation noise is known,
and in practice often achieves acceptable performance.} We choose not
to do so for the following reason: These models reflect the best
conjectures regarding the composition and structure of CEGP
atmospheres at this time, with little or no direct measurements of
their properties.  A matched filter approach based on these models
could potentially suffer from a loss in sensitivity should the actual
planetary atmospheres differ significantly from the current
assumptions.  On the other hand, the general shape and amplitude
predicted by the models are likely to be useful in gauging the
efficiency of the proposed detector.

\begin{figure}
\plottwo{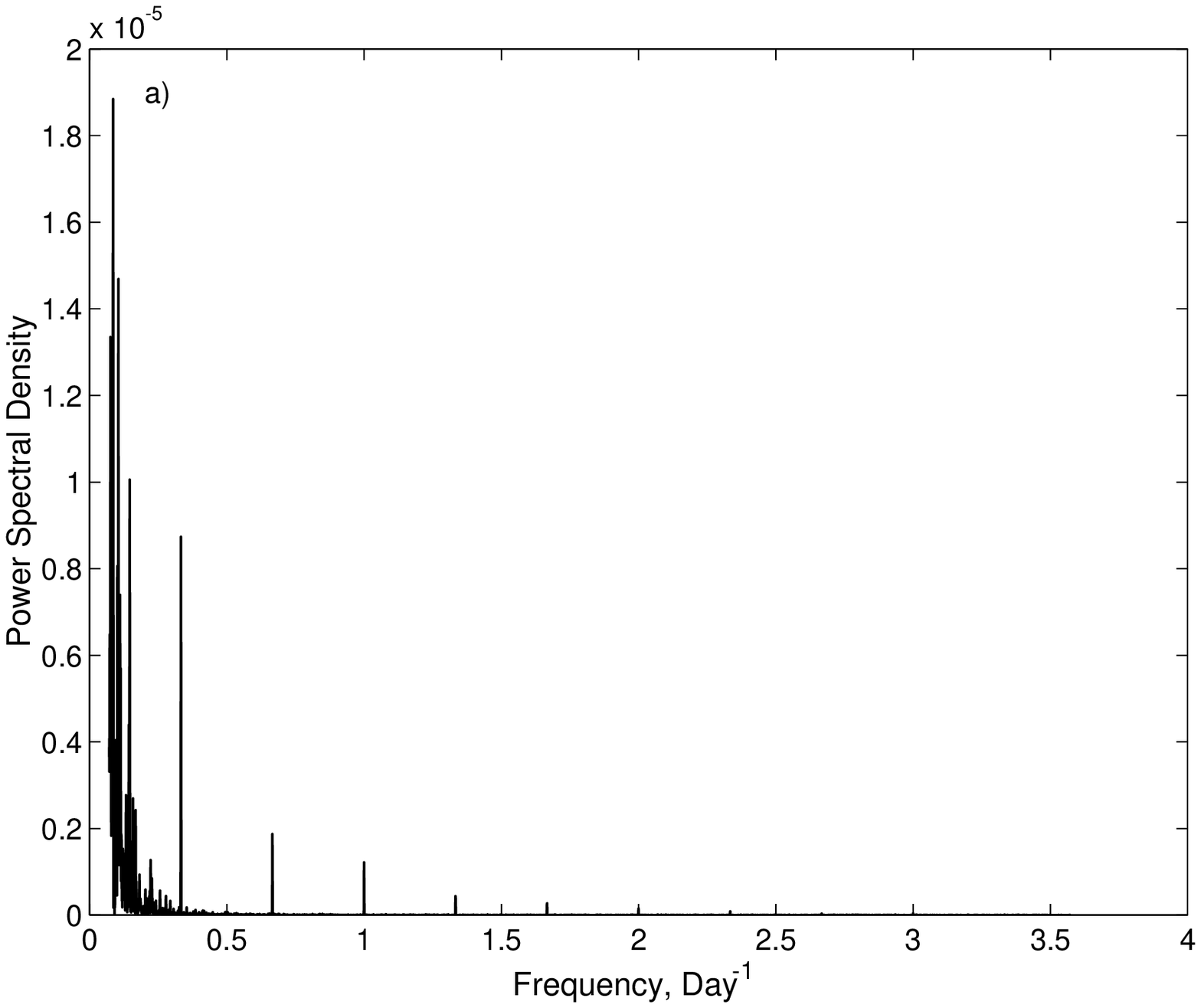}{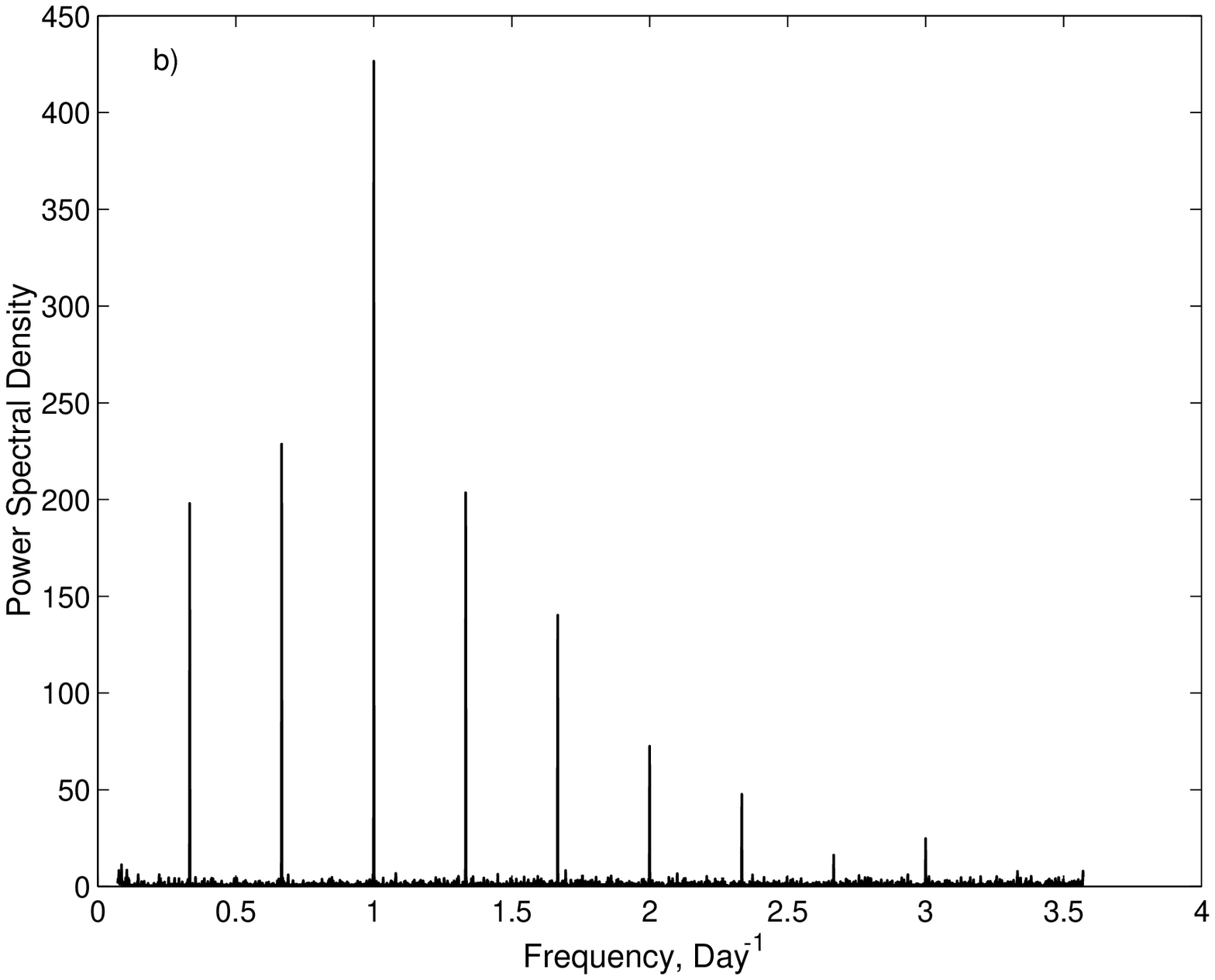} \caption[F2.eps]{The process of applying
the proposed detector to photometric data is illustrated by a) the
periodogram of synthetic stellar variability for a solar-like star
with a solar rotation period of 26.6 days, $\mr$=12 and an orbiting
1.2 R$_{\mathrm{J}}$ planet with an orbital period of 3 days, and b)
the ``whitened'' periodogram.  The components of the signal due to the
planet appear at multiples of 1/3 day$^{-1}$.  The fundamental is not
the strongest component in the whitened spectrum, as it would be for
the case of white observational noise.}
\label{fig:whitened}	
\end{figure}

\begin{figure}
\plotone{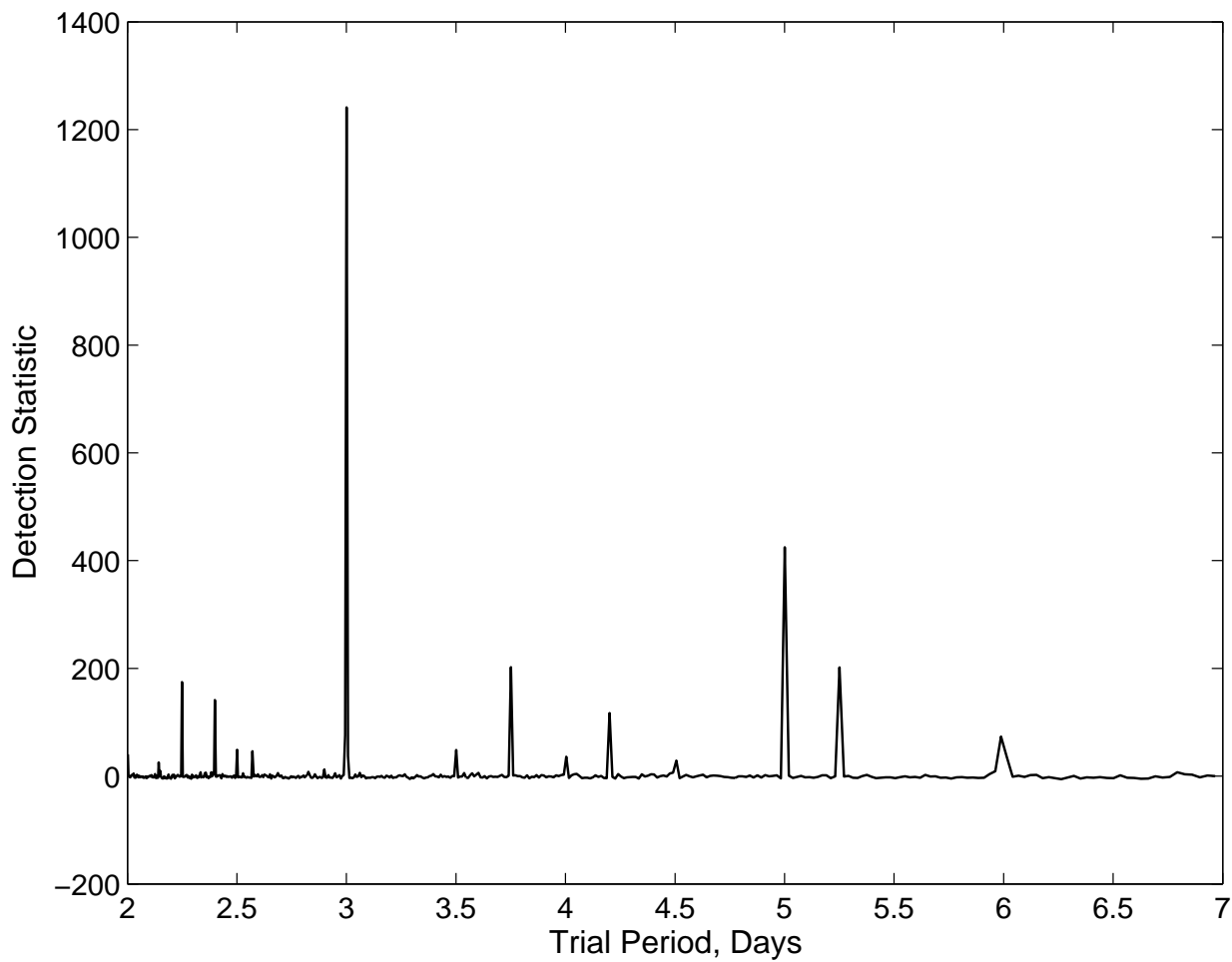} \caption[F3.eps]{The co-added spectrum corresponding
to the time series in Fig.  \ref{fig:whitened} is shown.  The
periodogram has been co-added to itself so that the components of a
periodic signal appear in the same bin, and thus, dramatically
increase the chance of detection.  Note the strong peak at 3 days,
corresponding to the period of the signal in the time series.  This
may not always be the case as it depends on the strength of the
fundamental compared to the background stellar and instrumental noise. 
In any case, the presence of many strong peaks at rational harmonics
of the actual fundamental provide additional confidence that a
periodic signal has been detected, and their spacing dictates the
fundamental period.  }
\label{fig:foldspec}	
\end{figure}

Our detector consists of taking the periodogram as an estimate of the
power spectral density (PSD) of the observations, estimating the
broadband background power spectrum of the measurement noise,
``whitening'' the PSD, and then forming detection statistics from the
whitened PSD. We first form a Hanning-windowed periodogram of the
$N$-point observations.  For convenience, we assume the number of
samples is a power of 2.  For \emph{Kepler's} sampling rate, $f_{s}$=4
hr$^{-1}$, $N=2^{17}$ points corresponds to 3.74 yr or about 4 yr. 
The broadband background, consisting of stellar variability and
instrumental noise, is estimated by first applying a 21-point moving
median filter (which replaces each point by the median of the 21
nearest points), followed by applying a 195-point moving average
filter (or boxcar filter).  The moving median filter tends to reject
outliers from its estimate of the average power level, preserving
signatures of coherent signals in the whitened PSD. The length of 195
points for the moving average corresponds to the number of frequency
bins between harmonics of a 7 day period planet for the assumed
sampling rate and length of the observations.  Both of these numbers
are somewhat arbitrary: wider filters reject more noise but don't
track the power spectrum as well as shorter filters do in regions
where the PSD is changing rapidly.  This background noise estimate is
divided into the periodogram point-wise, yielding a ``whitened''
spectrum as in Figure~\ref{fig:whitened}.  The advantage of whitening
the periodogram is that the statistical distribution of each frequency
bin is uniform for all frequencies except near the Nyquist frequency
and near DC (a frequency of 0), simplifying the task of establishing
appropriate detection thresholds.  The whitened periodogram is
adjusted to have an approximate mean of 1.0 by dividing it by a factor
of 0.6931, the median of a $\chi_{2}^{2}(2 x)$ process.  (This
adjustment is necessitated by the moving median filter.)  Finally, the
value 1 is subtracted to yield a zero-mean spectrum.  [The
distribution of the periodogram of zero-mean, unit-variance WGN is
$\chi_{2}^{2}(2 x)$ \citep[see, \eg,][]{papoulis84}.]  Finally, the
detection statistic for each trial period $N/\left(K f_{s}\right)$ is
formed by adding the bins with center frequencies $i K f_{s}/N$,
$i=1,\ldots,M$ together, where $M\le 7$, as in
Figure~\ref{fig:foldspec}.  The trial periods are constrained to be
inverses of the frequency bins between 1/2 and 1/7 days$^{-1}$.

\begin{figure}
\plotone{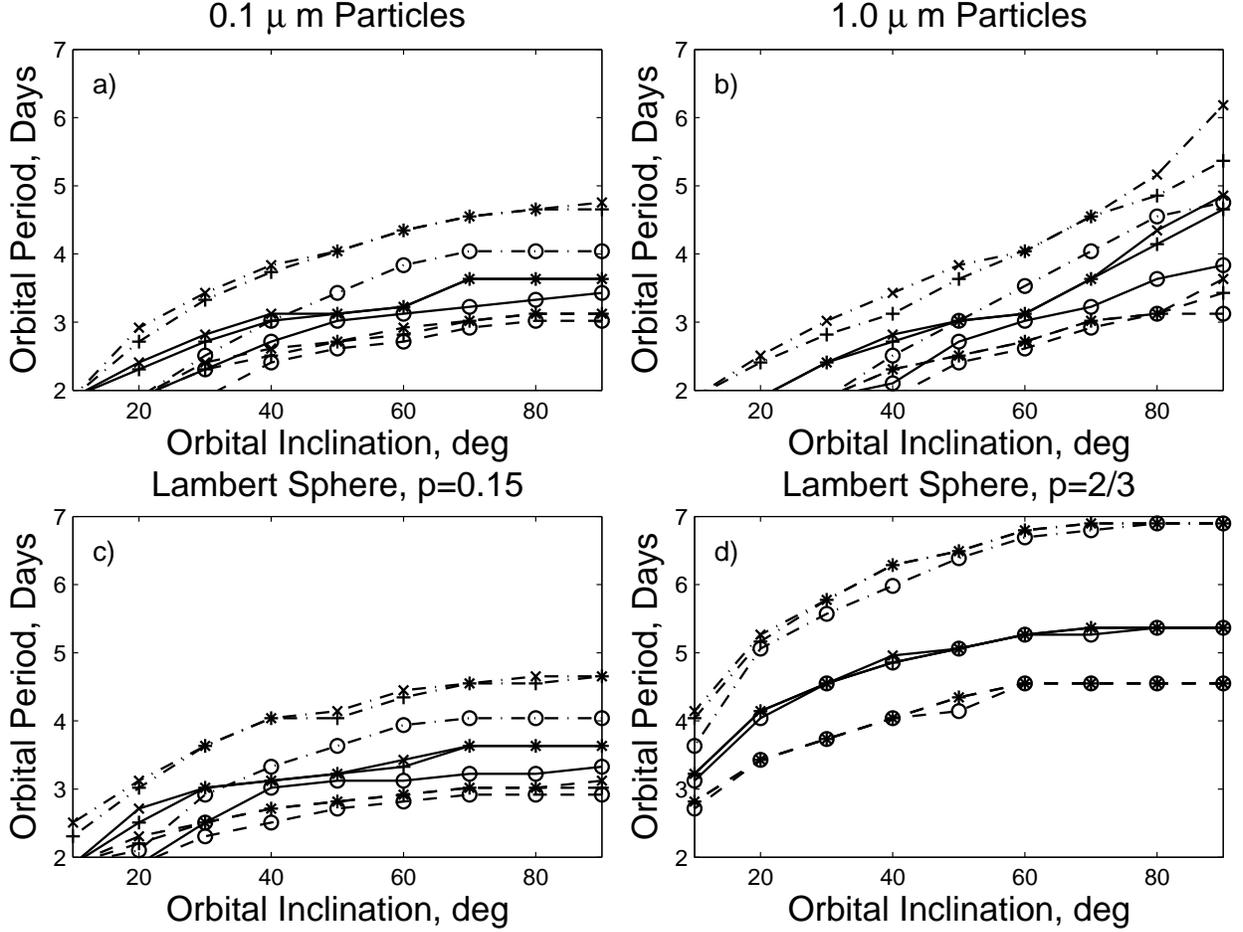} \caption[F4.eps]{The maximum detectable planetary
period at a detection rate of 90\% vs.  orbital inclination for
various stellar brightnesses and rotation periods and 4 yr of data are
plotted for: a) Seager \etal's $\bar{r}=0.1$ $\mu$m particle model, b)
Seager \etal's $\bar{r}=1.0$ $\mu$m particle model, c) a Lambert
sphere with geometric albedo $p=0.15$, and d) a Lambert sphere with
$p=2/3$.  The number of assumed Fourier components, $M$, is set to one
here.  Stellar rotation periods of 20 days, 25 days and 35 days are
denoted by dashed lines, solid lines and dash-dotted lines, respectively. 
Stellar magnitudes $\mr$=9.5, 11.5 and 13.5 are denoted by `x's,
crosses, and open circles, respectively.  The first three models yield
comparable numbers of expected CEGP detections.  Seager \etal's
$\bar{r}=1.0$ $\mu$m particle model is easier to detect at longer
periods at high orbital inclinations relative to the $\bar{r}=0.1$
$\mu$m particle model or the $p=0.15$ Lambert sphere model.  This is
due to the greater number of Fourier components, which can compensate
for red noise from stellar variability that can mask lower frequency
harmonics.}
\label{fig:Pmax}
\end{figure}

\placefigure{fig:whitened}
\placefigure{fig:foldspec}
\placefigure{fig:Pmax}

This procedure was applied to each of 450 model reflected light curves
spanning inclinations from 10$^{\circ}$ to 90$^{\circ}$, orbital
periods from 2 to 7 days, plus stellar variability for stars with
$P_{rot}$ between 5 and 40 days and instrumental and shot noise
corresponding to apparent stellar brightnesses between R=9.0 and
R=15.0.  The combinations of these parameters generated a total of
21,600 synthetic PSDs for which the corresponding detection statistics
were calculated.  The number of assumed Fourier components was varied
from $M=1$ to $M=7$.  Some results of these numerical trials are
summarized in Figure~\ref{fig:Pmax}, which plots the maximum
detectable orbital period, $P_{\mathrm{max}}$, for $M=1$ at a
detection rate of 90\% against $I$, for $P_{rot}$=20, 25 and 35 days,
for Sun-like (G2V) stars with apparent stellar magnitudes $\mr$=9.5,
11.5 and 13.5.  Detection thresholds and detection rates are discussed
in \S\ref{s:montecarlo}.

For $\bar{r}=0.1$ $\mu$m clouds (Fig.~\ref{fig:Pmax}a), planets are
detectable out to $P=4.75$ days for $P_{rot}$=35 days, out to $P=3.7$
days for $P_{rot}=25$ days, and out to $P=3.1$ days for $P_{rot}=20$
days.  The curves are rounded as they fall at lower inclinations, and
planets with $I$ as low as 50$^{\circ}$ are detectable for all the
curves, while planets with $I>20^{\circ}$ are detectable only for
stars with $P_{rot}=35$ days.  For clouds consisting of $\bar{r}=1.0$
$\mu$m particles (Fig.~\ref{fig:Pmax}b), the curves of
$P_{\mathrm{max}}$ are more linear, extending to orbital periods as
long as 6 days for $P_{rot}=35$ days, as long as 4.8 days for
$P_{rot}=25$ days, and to $>3$ days at high inclinations for stars
brighter than $\mr$=14.  The detectability of both of these models at
high orbital inclinations would be improved by searching for more than
one Fourier component, (\ie, choosing a higher value for $M$).  This
is a consequence of the larger number of harmonics in the reflected
light signature.  Although the power is distributed among more
components, as the orbital period increases, the signal is less
sensitive to the low frequency noise power due to stellar variability,
which easily masks the low frequency components of the signal.  The
behavior of the maximum detectable planetary radius for a Lambert
sphere with $p=0.15$ (Fig.  \ref{fig:Pmax}c) is very similar to Seager
\etal's $\bar{r}=0.1$ $\mu$m model.  A Lambert sphere with $p=2/3$
outperforms all the other models, as expected due to its significantly
more powerful signal.  Planets in orbits up to nearly 7 days can be
detected for Sun-like stars with rotation periods of 35 days.  For
Sun-like stars with rotation periods of 25 and 20 days, planets are
detectable with orbital periods up to 5.4 and 4.6 days, respectively. 
The Lambert sphere model PSD's contain only two Fourier components. 
Consequently, the detectability of such signatures is not improved
significantly by choosing $M>1$.

Now that we have specified the detector, we must analyze its
performance for the stellar population and expected planetary
population.  We should also determine the optimal number,
$M_{\mathrm{opt}}$, of Fourier components to search for, if possible. 
The value of doing so cannot be overstated: higher values of $M$
require higher detection thresholds to achieve a given false alarm
rate.  If too large a value for $M$ is chosen then adding additional
periodogram values for $M>M_{\mathrm{opt}}$ simply adds noise to the
detection statistic.  This will drive down the total number of
expected detections.  On the other hand, if too small a value for $M$
is chosen, then the sensitivity of the detector to CEGP signatures
would suffer and here, too, the number of expected detections would
not be maximized.  The first step is to determine the appropriate
threshold for the desired false alarm rate as a function of $M$.  This
is accomplished via Monte Carlo runs as presented in
\S\ref{s:montecarlo}.  To determine the best value of $M$, we also
need a model for the population of target stars, which defines the
observation noise, and a model for the distribution of CEGPs.  We use
the Besan\c{c}on galactic model to characterize the target star
population (\S\ref{s:besancon}).  The distribution of CEGPs with
orbital period can be estimated from the list of known CEGPs. 
Moreover, we need a method for extrapolating solar-like variability
from that of the Sun to the other spectral types.  Two methods are
considered and discussed in \S\ref{s:numdetections}.  In the first,
the stellar variability is treated strictly as a function of stellar
rotation period, so that the detection statistics are adjusted for the
varying stellar size.  In the second, it is assumed that the
mitigating effects of decreasing (increasing) the stellar area towards
cooler (warmer) late-type stars are exactly balanced by an increase
(decrease) in stellar variability.  Hence, no adjustment is made to
the detection statistics as a function of spectral type.  Given this
information, we can then determine which value of $M$ maximizes the
number of expected CEGP detections for a particular atmospheric model.

We found that the optimal value of $M$ depends a great deal on the
assumed stellar population, and the distribution of CEGPs with orbital
period.  If the rotation periods of \emph{Kepler's} target stars were
evenly distributed, then optimal values for $M$ varied from $M=1$ to
$5$, depending on the atmospheric model and method for extrapolating
stellar variability across spectral type.  Adopting a realistic
distribution of stellar rotation period and spectral type produced a
surprising result.  We found that $M=1$ yielded the highest number of
detections assuming all four of the atmospheric models considered were
equally likely.  The number of detections for each atmospheric model
as a function of $M$, and the average number of detections across all
four atmospheric models are given in Table \ref{ndetfouriertbl}.  The
results of both methods for extrapolating stellar variability across
spectral type are averaged together for this exercise.  The effects of
setting $M$ to 1 were not strong for Seager \etal's $\bar{r}$=1.0
$\mu$m model where $M_{\mathrm{opt}}$ exceeded 1.  In this case, $M=2$
or 3 was optimal, depending on how stellar variability was
extrapolated.  Up to 6\% fewer CEGPs would be detected using $M=1$
rather than $M=3$ (174 vs.  185 total detections).  For Seager \etal's
$\bar{r}=$0.1 $\mu$m model and both Lambert sphere models, $M=1$ was
optimal, although the average number of detections drops slowly with
$M$.

\placetable{ndetfouriertbl}


\section{Monte Carlo Analysis}
\label{s:montecarlo} 

In order to determine the detection thresholds and the corresponding
detection rates, we performed Monte Carlo experiments on WGN
sequences.  Much of this discussion draws on that of
\citet{jenkinsetal2002}, which concerns the analogous problem of
establishing thresholds for transit searches.  Each random time series
was subjected to the same whitening, and spectral co-adding as
described in \S\ref{s:detapproach}.  Two statistical distributions
produced by these Monte Carlo trials are of interest: that of the null
statistics for a single trial period, and that of the maximum null
statistic observed for a search over all the trial periods.  The
former defines in part the probability of detection for a given
planetary signature and background noise environment, since the
distribution of the detection statistic in the presence of a planet
can be approximated by shifting the null distribution by the mean
detection statistic.  The latter dictates the threshold necessary to
control the total number of false alarms for a search over a given
number of stars.

Let $l_{1,0}(M)$ denote the random process associated with the null
statistics for a single trial period, and assumed number of Fourier
components, $M$.  Likewise, let $l_{\mathrm{max},0}(M)$ denote the
random process corresponding to the null statistics for a search of a
single light curve over all trial periods.  The corresponding
cumulative distribution functions are $P_{l_{1,0}}(x,M)$ and
$P_{l_{\mathrm{max},0}}(x,M)$, respectively.\footnote{In this
discussion, the cumulative distribution function of a random variable
$y$ is defined as the probability that a sample will not exceed the
value $x$: $P_{y}(x)=P(y\le x)$.  The complementary distribution
function, $1-P_{y}(x)$ will be denoted as $Q_{y}(x)$.} For $N_{*}$
stars, the thresholds, $\eta(M)$, that yield a false alarm rate of
$1/N_{*}$ for each search are those values of $x$ for which
\begin{equation}
	Q_{\lmax}(x,M)=1-P_{\lmax}(x,M)=1-1/N_{*}
\label{eq:searchfa}
\end{equation}  
and hence, deliver a total expected number of false alarms of exactly
one for a search of $N_{*}$ light curves.  For a given threshold,
$\eta$, and mean detection statistic, $\bar{l}_{1}(M)$, corresponding
to a given planetary signature the detection rate, $P_{D}(M)$, is
given by
\begin{equation}
	P_{D}(M)=P_{l_{1,0}}(\bar{l}_{1}-\eta,M),
\label{eq:dr1}
\end{equation}
where the explicit dependence of $\bar{l}_{1}$ and $\eta$ on $M$ is
suppressed for clarity.

\begin{figure}
\plottwo{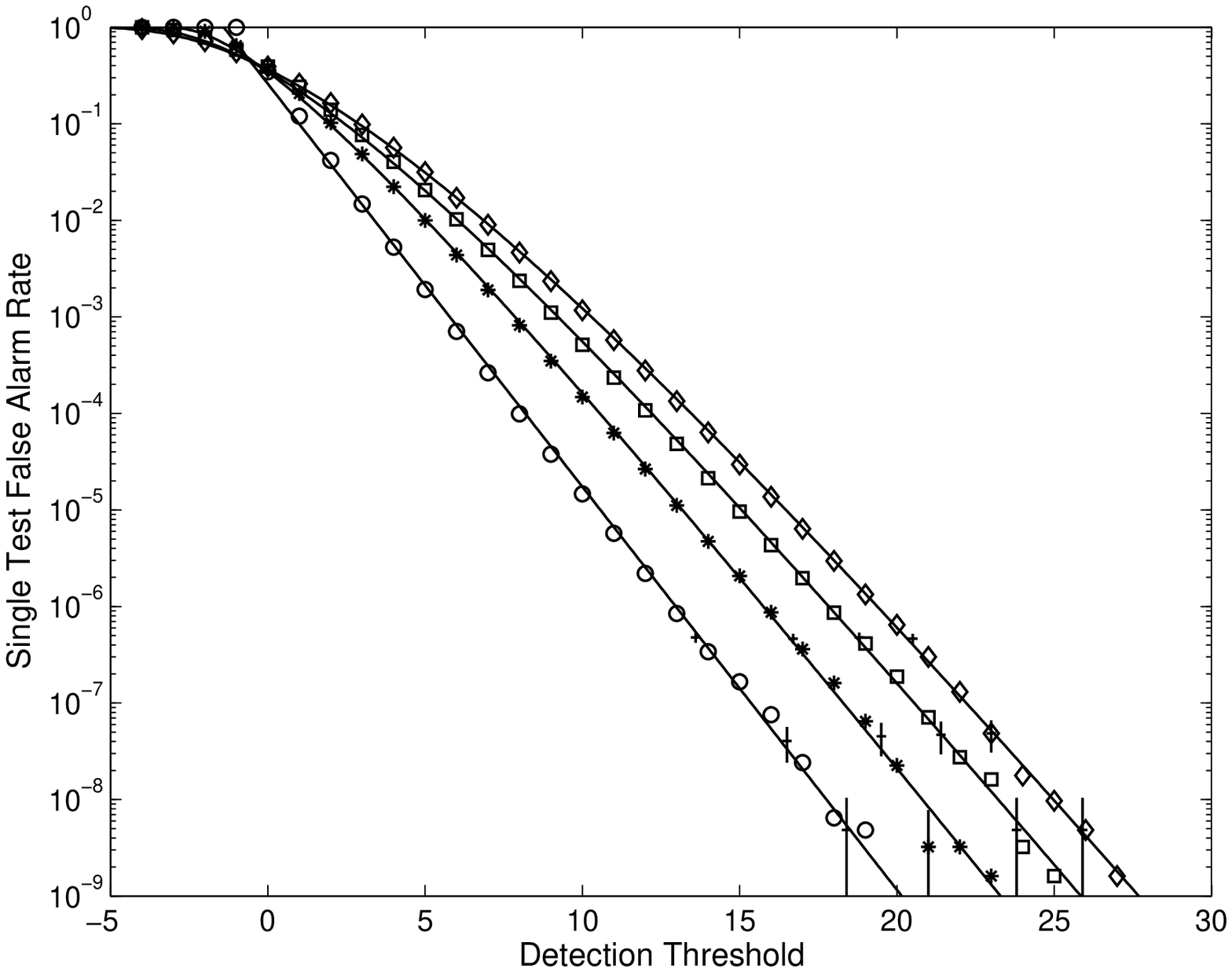}{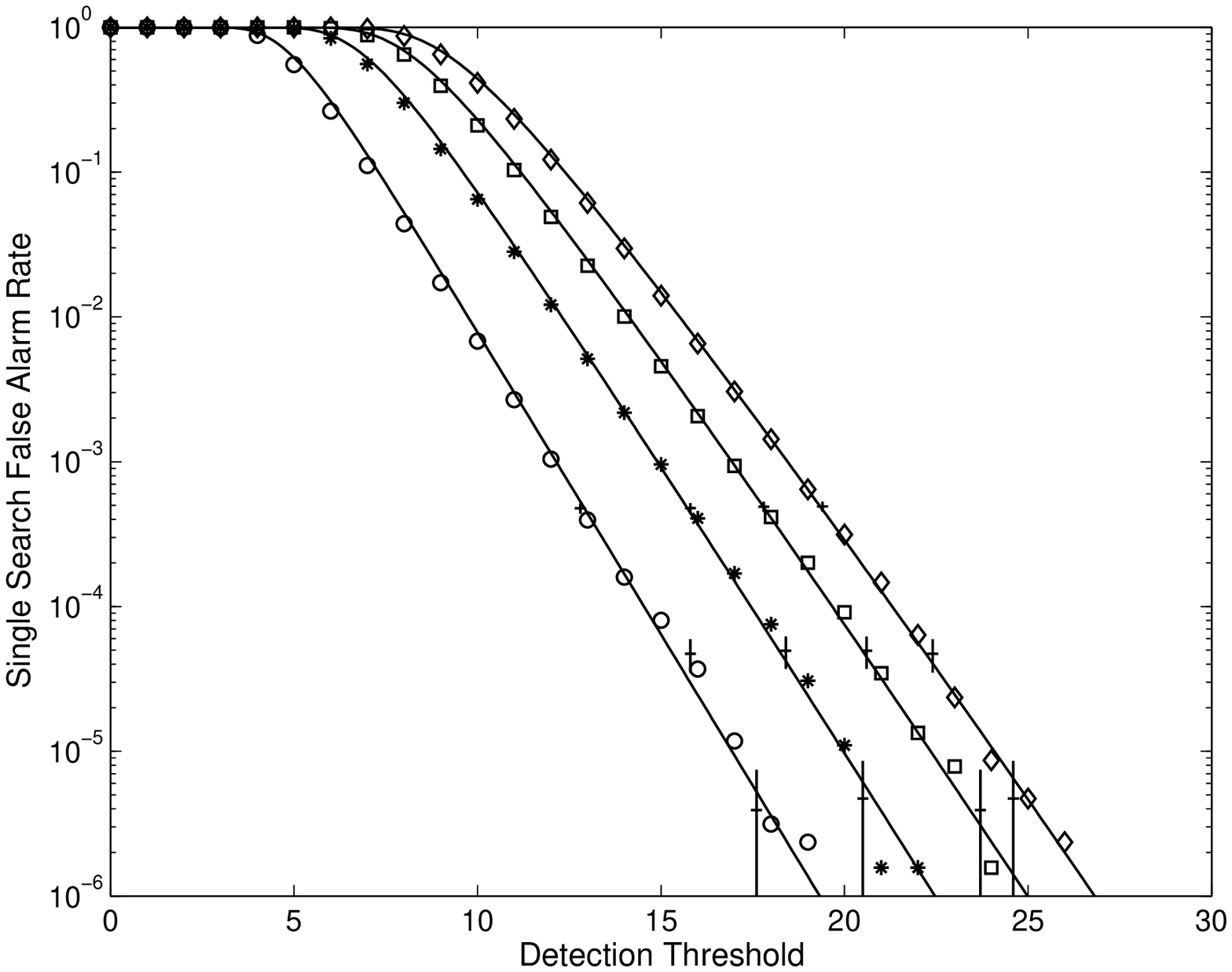} \caption[F5.eps]{The single test a) and
single search b) false alarm rates as functions of detection threshold
for the proposed detector.  The number of assumed Fourier components,
$M$=1, 3, 5 and 7, are denoted by circles, asterisks, squares, and
diamonds, respectively, for the sample distributions.  For clarity,
only every fifth point of each sample distribution is plotted.  The
solid curves indicate the least-squares fits to the log of the sample
distributions, emphasizing the upper tail in the fit.  Error bars for
95\% confidence intervals are denoted by vertical line segments
crossed by horizontal line segments at various locations in each
sample distribution.  The single test false alarm rates can be used to
estimate the detection rates for a given CEGP signal (see Fig. 
\ref{fig:singletestDR}), while the single search false alarm rates
determine the detection threshold for a given number of target stars
and desired total number of false alarms.  Determining the optimal
value of $M$ is important, given that higher values of $M$ require
correspondingly higher detection thresholds, which drives down the
number of detections if the chosen value of $M$ is too high.}
\label{fig:FAR}   
\end{figure}

\placefigure{fig:FAR}

Figure \ref{fig:FAR}a shows the sample distributions for
$Q_{\lone}(x,M)$ resulting from 619 million Monte Carlo trials for
$M=1$, 3, 5, and 7.  This represents the single test false alarm rate
as a function of detection threshold.  Figure \ref{fig:FAR}b shows
$Q_{\lmax}(x,M)$ resulting from 1.3 million Monte Carlo runs, for the
same values of $M$.  This represents the single search false alarm
rate as a function of detection threshold for each value of $M$. 
Error bars denoting the 95\% confidence intervals appear at selected
points in both panels.  

It is useful to model $\plone$ and $\qlmax$ analytically.  If the
whitening procedure were perfect, and assuming that the observation
noise were Gaussian (though not necessarily white), $\lone$ would be
distributed as a $\chi^{2}_{2\,M}$ random variable with a
corresponding distribution $Q_{\chi^{2}_{2\,M}}(2\,x+2\,M)$.  Figure
\ref{fig:FAR}a shows the sample distributions for $\lone$ resulting
from 619 million Monte Carlo runs.  Higher values of $M$ require
higher thresholds to achieve a given false alarm rate.  We fit
analytic functions of the form
\begin{equation}
Q_{\lone}(x,M)\approx Q_{\chi^{2}_{2\,M}}(A\,x+B)
\end{equation}
to the sample distributions $Q_{\lone}(x,M)$, where parameters $A$ and
$B$ allow for shifts and scalings of the underlying analytical
distributions.  Two methods for determining the fitted parameters are
considered.  In the first, we fit the analytic expressions directly to
the sample distributions, including the uncertainties in each
histogram bin.  The resulting fit is useful for estimating the
detection rate as a function of signal strength above the threshold,
but may not fit the tail of the distribution well.  In the second
method, the log of the analytic function is fitted to the log of the
sample distributions in order to emphasize the tail.  The fitted
parameters are given in Table \ref{nfittbl}.  Regardless of whether
the sample distribution or the log sample distribution is fitted, the
values for $A$ are within a few percent of 2 and the values of $B$ are
no more than 14\% different from $2\,M$, indicating good agreement
with the theoretical expectations.

\placetable{nfittbl}

To determine the appropriate detection thresholds, we need to examine
the sample distributions $\qlmax$.  These are likely to be
well-modeled as the result of taking the maximum of some number,
$N_{\mathrm{EIT}}$, of independent draws from scaled and shifted
$\chi^{2}_{2\,M}$ distributions.  Here, $N_{\mathrm{EIT}}$ is the
effective number of independent tests conducted in searching for
reflected light signatures of unknown period in a single light curve. 
We take the values for $A$ and $B$ obtained from the fits to the log
of $Q_{\lone}(x,M)$ and fit the log of the analytic functions of the
form
\begin{equation}
	Q_{\lmax(x,M)}\approx 1-P_{\chi^{2}_{2\,M}}^{N_{\mathrm{EIT}}}
	\left(A,x+ B\right)
\end{equation}
to the log of the sample distributions $Q_{\lmax}(x,M)$.  The values
for $N_{\mathrm{EIT}}$ are given in Table \ref{nfittbl} and fall
between 430 and 476.  For the length of data considered, there are
$\sim$490 frequency bins corresponding to periods between 2 and 7
days.  Thus the whitening and spectral co-adding operations apparently
introduce some correlation among the resulting detection statistics,
somewhat reducing the total number of independent tests conducted per
search.


In determining the expected number of CEGPs whose reflected light
signatures \emph{Kepler} will likely detect, we average the detection
rates from \S\ref{s:detapproach} over all inclinations and over the
distribution of planetary periods of known CEGPs (see
\S\ref{s:besancon}).  The former can be accomplished by noting that
inclination for randomly oriented orbits is distributed according to
the sine function.  Table \ref{avgDRtbl} contains the average
detection rates for 1.2 \rj planets orbiting Sun-like stars as
functions of stellar rotation period and apparent magnitude for all
four atmospheric models for a detector with $M=1$.  These results
correspond to a false alarm rate of 1 in $10^{5}$ light curve
searches.  The detection rate falls more rapidly with decreasing
stellar rotation period than it does with increasing apparent stellar
magnitude for the range of magnitudes and rotation periods considered
here.  The atmospheric models predicted by \citet{seageretal2000} are
sensitive to the planet-star separation and are not likely to be
accurate for planets well within 0.04 AU or planets much beyond 0.05
AU. Most of the planets making up our assumed planetary orbit
distribution function fall within or close to these limits.  Thus, we
do not believe that departures from the simple scaling suggested by
\citet{seageretal2000} are important in estimating the number of CEGPs
that \emph{Kepler} will detect.  The detection rate is zero for stars
with rotation periods shorter than 20 days for all save the $p=2/3$
Lambert sphere model which can detect planets orbiting stars with
$P_{rot}$ as short as 15 days.

\section{Expected Number of Detections}\label{s:numdetections}
In this section we use the results of \S\ref{s:besancon},
\S\ref{s:detapproach} and \S\ref{s:montecarlo} along with statistics
of the known CEGPs to estimate the expected number of detections of
CEGPs by reflected light for \emph{Kepler}.  As discussed in
\S\ref{s:detapproach}, the results depend on $M$, the assumed number
of harmonics in the CEGP light curve.  Here we discuss in detail only
the results obtained by setting $M=1$, which maximized the number of
detections assuming all four atmospheric models were equally likely. 
Throughout this discussion we assume a uniform radius of 1.2
R$_{\mathrm{J}}$ for all CEGPs.  This is somewhat conservative: the
only CEGP with a known radius is the celebrated case of HD209458b with
a radius of 1.4 R$_{\mathrm{J}}$ \citep{codysasselov2002}.

Two methods to extrapolate stellar variability as a function of
spectral type are considered.  In the first, we assume that the
relationship between rotation period and stellar variability holds for
all spectral types, so that we modify the detection rates by
accounting for the dependence on the signal amplitude with the area of
the star, $R_{*}^{-2}$.  That is, the PSD of stellar variability is
assumed to be a function of stellar rotation period alone, while the
amplitude of the planetary signature also depends on the size of the
star.  We denote this method as spectral type compensation method A.
The second method, B, starts with the general observation that warmer
late-type stars tend to be more photometrically quiescent than cooler
late-type stars at time scales relevant to detecting CEGPs and further
assumes that this relationship compensates exactly for the dependence
of signal strength on stellar radius.  That is, for this model, no
adjustment for stellar radius is made to the amplitude of the
planetary signature.  The validity of these approaches will be tested
by \emph{Kepler} and the other upcoming photometry space missions.  An
important point is that the proposed detector can be tuned to the
actual observation noise via Monte Carlo techniques since CEGPs are
relatively rare.  The two approaches considered here are not expected
to be 100\% valid, but should produce reliable estimates for the
number of CEGPs \emph{Kepler} can expect to detect.

\placetable{nrefltbla}
\placetable{nrefltblb}

The expected number of CEGPs detected under spectral type compensation
method A are given in Table \ref{nrefltbla}, which lists the number of
detected planets for each of the four planetary atmospheric models. 
Seager \etal's $\bar{r}$=1.0 $\mu$m and $\bar{r}$=0.1 $\mu$m models,
and the Lambert sphere model with $p=0.15$ all detect about the same
number of planets: $\sim$120 to $\mr=14.0$ and $\sim$220 to
$\mr$=15.0.  For G-type stars, $\sim$65 CEGPs to $\mr$=14.0 and
$\sim$114 CEGPs to $\mr$=15.0 should be detected for each of these
models.  Among the four atmospheric models and spectral type
compensation method A, the Lambert sphere model with $p=2/3$ stands
out.  In this case, 230 and 712 CEGPs would be detected to $\mr$=14.0,
and $\mr$=15.0, respectively.  A little over half of these detections
would occur around G-type stars.  In contrast to the other atmospheric
models, significant numbers of CEGPs would be detected around F-type
stars, in addition to G-, K- and M-type stars.

Table \ref{nrefltblb} contains the estimates for the number of CEGPs
detected under spectral type compensation method B. Here, for Seager
\etal's models and for the $p=0.15$ Lambert sphere, the number of
expected detections drop by about $\sim$40\% to $\mr$=14.0 and by
$\sim$50\% to $\mr$=15.0.  About 80 planets are detected to $\mr$=14.0
($\sim$55 orbiting G-type stars), and $\sim$115 planets are detected
to $\mr$=15.0 ($\sim$80 orbiting G-type stars).  In contrast, the
$p=2/3$ Lambert sphere model detects $\sim$8\% more planets than it
does for method A. The reason for the differences between method A and
method B is that the detections are shifted to earlier type stars. 
The $p=2/3$ Lambert sphere model gains detections because there are
more F-type target stars than G-type target stars according to the
galactic model, which offsets the lower detectability of the CEGP
signal for the faster rotating F stars.  The detectabilities of the
CEGP signals for the other atmospheric models are more sharply reduced
for F stars, so that while the detections shift towards F stars, an
insufficient number are gained to offset the reduced number of
detections for later star types.

As expected, more detections are obtained if stars as dim as
$\mr$=15.0 are observed.  There are 132,000 more stars between
$\mr$=14.0 and $\mr$=15.0 than there are to $\mr$=14.0 ($\sim$80,000),
for a total of $\sim$200,000 main-sequence stars in the FOV.
\emph{Kepler's} downlink and onboard storage system are capable of
handling $\sim$200,000 target stars for the first year of operation,
and $\sim$140,000 thereafter.  The most promising targets are those
stars with long rotation periods.  This can be ascertained after
several months of observation once a good PSD estimate can be obtained
and the detection rates can be estimated.  It is likely that the
$\sim$200,000 main sequence stars to $\mr$=15.0 in \emph{Kepler's} FOV
can be pared down to $\sim$140,000 during the first year of
observation.  Thus, \emph{Kepler} can be expected to detect the
reflected light signatures of between 100 and 760 CEGPs (see Tables
\ref{nrefltbla} and \ref{nrefltblb}).  Note that the detections should
be evenly distributed in time: the energy of a reflected light
signature is directly proportional to the length of its time series. 
Thus, approximately 25\% of the discoveries, or between 25 and 190
CEGPs should be detected by reflected light within the first year of
the \emph{Kepler Mission}.

The highest detection rates occur for edge-on planetary orbits: those
most likely to produce transits.  Between 14\% and 40\% of the CEGPs
detected by reflected light will exhibit transits, depending on the
assumed atmospheric model and stellar variability model.  These
planets present an opportunity to extract the shape of the occultation
of the planet by its star.  In this case, the average brightness or DC
level of the reflected light signature can also be determined, which
is not the case for non-transiting CEGPs.  Moreover, since the
transits will almost certainly be detected in the first several weeks
of the mission, the requisite thresholds for detecting the reflected
light signatures can be significantly reduced, since the period is
constrained by the observations of the transits.  Given that $\sim$181
CEGPs in the FOV will exhibit transits, we should be able to constrain
if not measure the albedos quite well.

While solar variability may certainly not be safely extrapolated to
significantly different stellar classes, detections of CEGPs might
also be possible around G3 through early K giants because of their
expected low rotation rates, if the rotation period criteria in this
study is found to be applicable to giant stars.  M giants might be too
large to allow very close planets \citep[see, \eg,] [] {gray1992},
however K giants, because of their increased mass, would allow planets
50\% more distant with the same orbital periods discussed above.

\section{Potential Sources of Confusion and Methods of Discrimination}
\label{s:confusionanddiscrimination}

Detection algorithms detect all signals of sufficient amplitude with
features that are well matched to the shape of the signal of
interest.\footnote{An exception to this rule is provided by the
incoherent matched filter or ``energy detector'' that thresholds the
variance of a time series.  This detector is not sensitive to the
shape of the input signal, and consequently, suffers inferior
performance relative to a matched filter when the shape of the target
signal is well defined \citep[see, \eg,][]{kay1998}.} Thus, not all
signals yielding detection statistics above the detection threshold
need be signatures of CEGPs.  Indeed, several potential sources of
confusion exist that might inject signals similar to reflected light
signatures of CEGPs.  These include intrinsic photometric variability
of target stars themselves, and dim background variable stars within
the photometric apertures of target stars.  Such variations include
those produced by star spots, eclipsing or grazing eclipsing binaries,
or intrinsic stellar pulsations.  Section \S\ref{s:falsepositives}
describes each of these classes of variability along with an
assessment of the likelihood they pose as sources of confusion. 
Section \S\ref{s:centroiding} presents a robust method for rejecting
confusion from blended, variable background stars in a target star's
photometric aperture.

\subsection{Potential Sources of Confusion}
\label{s:falsepositives}

Sources of stellar variability that might be mistaken for reflected
light signatures of CEGPs include stellar pulsations, star spots, and
photometric variability induced by binarity.  These phenomena can
occur in the target star or in a blended background star, but the
amplitudes of concern are different since the magnitude of the
variations of a blended background star will be diluted by the flux of
the target star.  In addition, non-reflected light signatures of CEGPs
might be present, confounding the isolation and detection of the
reflected light signature.  In this section we discuss these sources
of photometric variability and assess the likelihood that each poses
as a source of confusion.

CEGPs can induce periodic photometric variations other than that due
to reflected light.  Doppler modulation of the host stellar spectrum
via reflex motion of the host star about the system barycenter
modulates the total flux observed in the photometer's bandpass. 
\citet{loebgaudi2003} estimate the amplitude of this effect and
conclude that Doppler-induced photometric variations for Jupiter-mass
planets orbiting solar-type stars in periods less than 7 days are
about 20 times fainter than the reflected light signature of
Jupiter-sized, $p=2/3$ Lambert spheres.  The Doppler-induced
photometric signal is 90$^{\circ}$ out of phase with that of the
reflected light component from a CEGP. Hence, rather than making it
more difficult to detect a CEGP, the combination of the two signatures
makes it easier to detect one since the power from orthogonal signals
add constructively in the frequency domain. Radial velocity 
measurements should help distinguish between the two signatures in 
the case of non-transiting CEGPs.

Stellar pulsations can cause strictly periodic photometric variations. 
Acoustic waves traveling in the Sun resonate at specific frequencies
with characteristic periods on the order of 5 minutes and typical
amplitudes of $\sim$10 ppm.  The coherence lifetime for these
so-called p-mode oscillations is approximately a month, beyond which
the sinusoidal components drift out of phase \citep{deubner1984}. 
Buoyancy waves (also called gravity waves) should have much longer
periods of 0.28-2.8 hours along with correspondingly longer coherence
timescales.  To date, no one has observed the signatures of g-modes in
the Sun.  The VIRGO experiment aboard \emph{SOHO} has placed upper
limits of 0.5 ppm on the amplitudes of solar g-modes
\citep{appourchaux2000}, which is in line with theoretical predictions
\citep{andersen1996}.  It does not appear that pulsations of
solar-like stars could present major problems: the coherence
timescales are short and the amplitudes are significantly smaller than
those due to the reflected light component from CEGPs.  Moreover, the
amplitudes preclude stellar pulsations of background blended stars
from being confused with signatures of CEGPs due to dilution.

Long-lived star spots or groups of spots can produce quasi-sinusoidal
photometric signatures.  Some individual starspot groups of F, G, and
K dwarfs have been known to last for months-to-years and cover an
appreciable fraction of the star's surface \citep[20-40\% in extreme
cases,][]{cram1989}, with the starspot cycles themselves lasting from
a half to several decades for nearby solar-type stars
\citep{baliunas1985}.  Contributions to solar variability at tens of
minutes come from granulation and are present in only a few tens of
ppm, while sunspots contribute a variation of about 0.2$\%$ over days
or weeks.  Faculae can also contribute variations of about $0.1\%$
over tens of days and last longer than individual sunspots, because
differential rotation distributes these over the whole solar disc
\citep{hudson1988}.  It is difficult to imagine that star spots on
solar-like single stars could be easily confused with CEGPs.  On the
Sun, for example, individual sunspots evolve and change continuously
on timescales comparable to the mean solar rotation period (26.6
days).  Thus, the photometric signatures of sunspots vary from
rotation to rotation so that the photometric dips due to spots do not
repeat with a great degree of precision.  In the Fourier domain it can
be difficult to identify the fundamental associated with the solar
rotation period: the peak is extremely broad.  Of more concern, then,
are photometric variations from dim background late-type binaries,
such as BY Dra or RS CVn variables.

The BY Draconis variables are dKe and dMe stars with typical
differential amplitudes of 0.2 magnitudes and periods of a few days. 
For example, in photometric observations of CM Draconis (M4 + M4, 1.27
day period), \citet{lacy1977} noted a $\sim$0.01 mag sinusoidal
feature he attributed to a long-lived, high latitude spot group that
persisted for years.  RS CVn stars are generally eclipsing binaries
consisting of at least one subgiant component.  These stars display
nearly sinusoidal variations of up to 0.6 mag.  The photometric
variations are due to an uneven distribution of cool spots in
longitude that rotationally modulate the apparent flux.  Fortunately,
one way of distinguishing these variations from the phase variations
of CEGPs is the fact that starspot activity of these stars varies with
phase over time.  \citet{kozhevnikov2000} found that the
quasi-sinusoidal starspot variation of CM Draconis had shifted by 60
degrees in phase over a two decade period and had increased in
amplitude (to $\sim$0.02 mag).  The eponymous BY Dra (M0 Ve + M0 Ve)
has a mean photometric period of 3.836 days, and can demonstrate
rather fickle photometric behavior: the nearly sinusoidal variations
discovered by \citet{chugainov1973} nearly disappeared by mid-1973. 
The light curves for several BY Dra and RS CVn stars can be explained
by the presence of two large spots on one of the stellar components. 
As the spots evolve and migrate in longitude, the photometric
variations change significantly \citep[see, \eg,][]{rodonoetal1986}. 
Some RS CVn systems with orbital rotation periods of several days
exhibit remarkable photometric variations over timescales of months. 
The RS CVn binary V711 Tau (K0 V + B5 V), for example, has an orbital
period of 2.84 days, and migration of spot groups in longitude leads
to changes in its ``photometric wave'' including the exhibition of
double peaks, nearly sinusoidal variations, and rather flat episodes
\citep{bartolinietal1983}.  Starspot-induced variations do not seem
likely candidates for being mistaken for reflected light signatures of
CEGPs, even for binary systems.

Ellipsoidal variables [\eg, o Persei (B1 III + B2 III), period = 4.42
days, differential amplitude 0.07 magnitudes in V] are non-eclipsing
binaries that display photometric variations due to the changing
rotational aspect of their tidally elongated shapes
\citep{sterkenjaschek1996}.  These stars' light curves exhibit two
maxima and two minima per orbital period, and one minimum can actually
be significantly deeper than the other.  Thus, we do not expect that
ellipsoidal variables will be mistaken for CEGPs as the shape of the
variations is significantly different from that expected for CEGPs.

It is unlikely that photometric variations of binary target stars will
be confused with CEGPs.  The \emph{Kepler Mission} will be preceded by
ground-based observations to characterize all the stars in the FOV
with $\mr\le16$.  These observations should be able to detect almost
all of the short period binaries.  Moreover, ground-based, follow-up
observations should be able to detect any of these types of variable
stars in the cases where one might have been mistakenly classified. 
These follow-up observations should help discriminate between
planetary and stellar sources for any candidate signatures of CEGPs. 
Nevertheless, we should examine the frequency of such binary systems
in the photometric apertures of target stars, and \emph{Kepler's}
ability to distinguish between photometric variability intrinsic to a
target and that due to blended background variables.

In a study of the light curves of 46,000 stars in the cluster 47 Tuc,
\citet{albrowetal2001} identified 71 likely BY Dra stars that
exhibited photometric variations as high as 0.2 magnitudes.  The
fraction of stars that are in binary systems is significantly lower in
47 Tuc ($\sim$14\%) than it is in the galactic disc \citep[$\sim$65\%,
as per][]{duquennoymayor1991}.  The peak-to-peak amplitudes of the
CEGP reflected light curves considered here are between 20 and 60 ppm,
so that background BY Dra binaries would need to be $\sim$8 magnitudes
dimmer than a particular target star to exhibit photometric variations
of the appropriate amplitude.  We determined the distribution of
late-type (G, K and M) stars with $\mr$=17.0 to 23.0 corresponding to
the range of apparent magnitudes for \emph{Kepler} target stars
discussed in \S\ref{s:numdetections} using the Besan\c{c}on galactic
model.  The number of binary systems with rotation periods between 2
and 7 days can be estimated using the Gaussian model of
\citet{duquennoymayor1991} for the distribution of binaries as a
function of the log period.  According to this distribution,
$\sim$1.75\% of binaries in the galactic disc should have periods in
this range.  Table \ref{nbackbintbl} gives the number of background
binaries with periods in this range consisting of at least one dwarf
G, K or M star in each aperture of a \emph{Kepler} target star.  The
apertures vary from 400 square arcsec for $\mr$=9.5 stars, to 200
square arcsec for $\mr=$14.5 stars, with a corresponding number of
background binaries varying from 13 to 69, respectively.  Even if such
a system appears in the photometric aperture of a target star, it is
likely that it can be detected by observing the centroid of the
brightness distribution over time (Ron Gilliland 2001, personal
communication), as discussed in \S\ref{s:centroiding}.

\placetable{nbackbintbl}

\subsection{A Method to Mitigate Confusion from Blended Background
Stars}\label{s:centroiding}

Since \emph{Kepler} will return target star pixels rather than stellar
fluxes to the ground, it will be possible to construct centroid time
series for all the target stars.  This represents a robust and
reliable means to discriminate between sources of variability
intrinsic to a target star and those due to background variable stars
situated within the target stars' photometric aperture.  Suppose that
the background variable located at $\bf{x_{2}}$ is separated from the
target star located at $\bf{x_{1}}$ by $\mathbf{\Delta
x}=\mathbf{x_{2}}-\mathbf{x_{1}}$, and that its brightness changes by
$\delta b_{2}$ from a mean brightness of $\bar b_{2}$, while the
target star's mean brightness is $\bar b_{1}$.  Then the change in the
photometric centroid position $\mathbf{\delta x_{c}}$ with respect to
the mean position is given by
\begin{equation}
	\mathbf{\delta x_{c}} = \delta b_{2} \; \mathbf{\Delta
	x}/(1+\bar b_{1}/\bar b_{2}).
\label{eq:centroid}
\end{equation}

Thus, a background star 8 magnitudes dimmer than the target star
separated by 1 arcsec and exhibiting a change in brightness of 10\%
will cause the measured centroid to change by 63 $\mu$as.  The
uncertainty in the centroid, however, is determined largely by the
Poisson statistics of the stellar flux signal and the random noise in
each pixel.  For \emph{Kepler's} Point Spread Function (PSF), the
uncertainty of the centroid of an $\mr$=9.5 star measured over a 24 hr
interval is $\sim$16 $\mu$as (on a single axis).  At a magnitude of
$\mr$=13.5, the corresponding uncertainty is $\sim$118 $\mu$as.  Note,
however, that we are not limited to the resolution of a centroid over
a short interval: Equation \ref{eq:centroid} implies that the time
series of the displacements of the target star's centroid will be
highly correlated with the photometric variations if the latter are
caused by a variable background star offset sufficiently from the
target star.  This fact implies that the centroid time series of a
star can be subjected to a periodogram-based test to determine if
there are statistically significant components at the photometric
period.  We performed numerical experiments with the PSF for
\emph{Kepler} and the expected shot and instrumental noise to
determine the radius to which background variables can be rejected at
a confidence level of 99.9\% for four years of observation.  The
expected accuracy of the centroids given above assumes that errors in
pointing can be removed perfectly by generating an astrometric grid
solution for \emph{Kepler's} target stars.  At some magnitude,
systematic errors will become significant.  Here, we assume that the
limiting radius inside which we cannot reject false positives is 1/8
pixels, or 0.5 arcsec.  Better isolation of background binaries might
be obtained in practice for stars brighter than $\mr=14.0$.  The
relevant figures for these calculations are given in Table
\ref{nbackbintbl}, showing that \emph{Kepler} should be able to reject
almost all such false positives for $\mr<$14.0.  A significant number
(28) of false positives might occur for target stars with
$14.0<\mr<15.0$.  These would require further follow-up observations
to help discriminate between background variables and signatures of
CEGPs.  We note, however, that this assumes that the background
variables display periodic signatures that retain coherence over
several years.  As discussed in \S\ref{s:falsepositives}, this is
generally not the case.

In summary, we do not expect intrinsic stellar variations to mimic
signatures of CEGPs over timescales of years.  The \emph{Kepler
Mission} incorporates a robust set of ground-based, follow-up
observations that include radial velocity studies as well as CaII H\&K
emission-line studies that can confirm starspot periodicities.  The
Doppler signatures of any candidate planets obtained by reflected
light can also be assessed by radial velocity measurements, a
relatively easy task for those stars with $\mr\le12$.  We note that
between 14\% and 40\% of the CEGPs detected by reflected light will
also exhibit transits, which together with the reflected light
signatures will provide another means of confirming many of the
candidates.

\section{Conclusions}\label{s:conclusions}
Although tailored for seeking Earth-sized planets via transit
photometry, NASA's \emph{Kepler Mission} is well positioned to detect
from 100 to $\sim$760 close-in giant inner planets (CEGPs) by
reflected light, depending on the presence of clouds and their
structure and composition.  The detector used in this analysis has a
threshold designed to produce no more than one false alarm for the
entire campaign.  Further, a combination of analysis of the candidate
stars' centroids and follow-up observations should reject most false
positives due to the injection of quasi-sinusoidal variations into the
target stars' apertures by variable stars.  For a given atmosphere,
the detectability is most sensitive to the stellar rotation period,
although stellar magnitude becomes important for $\mr>$12.5.  We can
state that it should be possible to discriminate between an atmosphere
composed of $\bar{r}$=0.1 $\mu$m clouds and one composed of
$\bar{r}$=1.0 $\mu$m clouds at high orbital inclinations, given the
great difference between the Fourier expansions of the predicted light
curves.  Both of Seager \etal's $\bar{r}$=1.0 $\mu$m and $\bar{r}$=0.1
$\mu$m cloud models, and a $p=0.15$ Lambert sphere model yield
comparable number of detections: $\sim$120 to $\mr$=14, and $\sim$220
to $\mr$=15.  If CEGP atmospheres are better characterized as $p=2/3$
Lambert spheres, then $\sim$250 and $\sim$760 CEGPS will be detected
to $\mr$=14, and to $\mr$=15, respectively.  This analysis is based on
realistic, yet preliminary models for CEGP atmospheres, as well as a
simple stellar variability model extrapolated from high precision
photometric observations of the Sun.  Clearly, the various near-term,
microsatellite photometry missions will permit development of better
models of stellar variability for a variety of main-sequence stars. 
Further, observations of stars with known CEGPs by these missions may
stimulate development of more comprehensive atmospheric models. 
Future work should incorporate emerging theories of CEGP atmospheres
and space-based photometry of a wider range of spectral types and
luminosity classes as they become available.  It should also address
the inverse problem of determining atmospheric parameters from
reflected light curves reconstructed from synthetic observations. 
This exercise may also better constrain the structure of the Fourier
series of such light curves, permitting the design of better detection
algorithms.

\acknowledgments We thank Ron Gilliland for helpful discussions
regarding the rejection of false positives through centroiding, and
for a careful reading of the paper.  We are grateful to Sara Seager
for making useful suggestions for expanding and improving the original
manuscript, and for helpful discussions regarding her models for the
atmospheres of CEGPs.  We thank Ren\'{e}e Schell for her detailed
editorial comments, and Doug Caldwell for a thorough reading.  We
acknowledge the efforts of the VIRGO team.  VIRGO is an investigation
on the solar and heliospheric observatory \emph{SOHO}, which is a
mission of international cooperation between ESA and NASA. LRD was
supported by the Carl Sagan Chair at the Center for the Study of Life
in the Universe, a division of the SETI Institute.  JMJ received
support from the \emph{Kepler Mission} Photometer and Science Office
at NASA Ames Research Center.  \emph{Kepler} is the 10th deep space
mission selected by NASA's Discovery Program.



\clearpage

\begin{figure}
\plotone{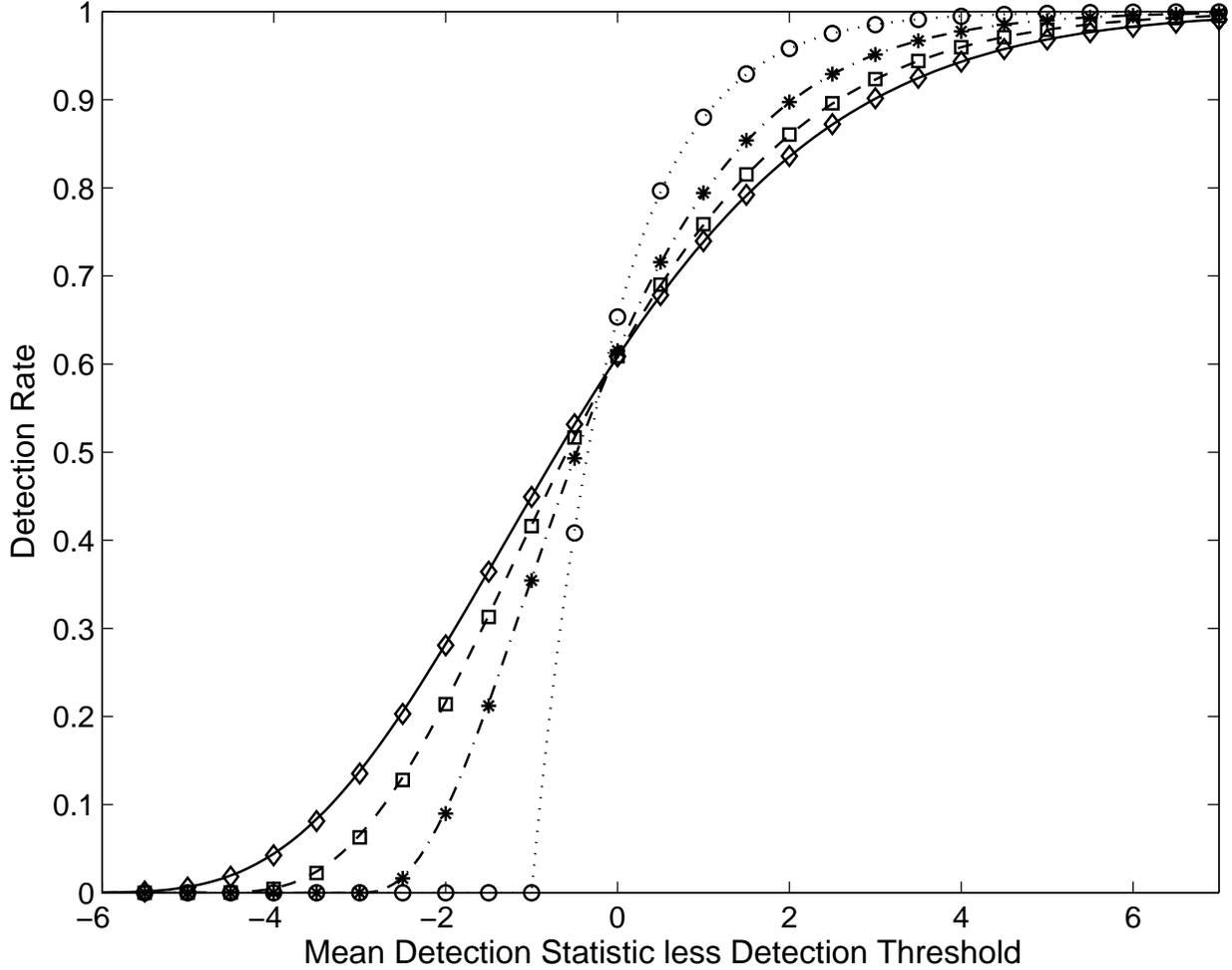} \caption[F6.eps]{The detection rate as a function of
the signal strength above the detection threshold (various symbols)
along with analytic expressions (various curves) fitted to the
empirical distributions.  The number of assumed Fourier components,
$M$=1, 3, 5 and 7, are denoted by circles, asterisks, squares and
diamonds, respectively for the sample distributions.  The
corresponding analytical fits are denoted by dotted, dash-dotted,
dashed and solid curves, respectively.  For clarity, only every 5th
point is plotted for the sample distributions.  At the threshold, the
detection rate attains $\sim$60\%.  This is due to the asymmetry of
the distribution of null statistics.  On this scale, the empirical
distribution functions and the analytic expressions appear identical.}
\label{fig:singletestDR}   
\end{figure}

\clearpage

\begin{deluxetable}{rrrrrrrr}
\tabletypesize{\small}
\tablecolumns{8}
\tablecaption{Modeled Number of Main-Sequence Stars in \emph{Kepler's}
Field of View
\label{nstartbl}}
\tablewidth{0pt} \tablehead{ 
\colhead{}&\multicolumn{6}{c}{Spectral Type}&\colhead{}\\
\colhead{$\mr$} & \colhead{B5} & \colhead{A5}
& \colhead{F5} & \colhead{G5} & \colhead{K5} & \colhead{M5} &
\colhead{All}} 
\startdata
  9.5&    151&    299&    200&     86&     20&      0&    756\\
 10.5&    481&    838&    706&    358&     80&      0&   2463\\
 11.5&   1002&   2181&   2248&   1300&    242&      7&   6979\\
 12.5&   1832&   5004&   7037&   4189&    991&     46&  19098\\
 13.5&   3051&  10245&  19796&  13379&   3271&    167&  49909\\
 14.5&   4498&  18142&  51098&  42035&  10969&    611& 127352\\
Total&  11014&  36708&  81085&  61347&  15573&    831& 206558\\
\enddata
\end{deluxetable}

\begin{deluxetable}{rrrrrrrr}
\tabletypesize{\small}
\tablecaption{Expected Number of Close-in Extrasolar Giant Planets
(CEGPs) in \emph{Kepler's} Field of View
\label{nCEGPtbl}}
\tablewidth{0pt} \tablehead{ 
\colhead{}&\multicolumn{6}{c}{Spectral Type}&\colhead{}\\
\colhead{$\mr$} & \colhead{B5} & \colhead{A5}
& \colhead{F5} & \colhead{G5} & \colhead{K5} & \colhead{M5} & \colhead{All} }
\startdata 
  9.5&      1&      3&      2&      1&      0&      0&      7\\
 10.5&      4&      7&      6&      3&      1&      0&     22\\
 11.5&      9&     19&     20&     11&      2&      0&     61\\
 12.5&     16&     44&     62&     37&      9&      0&    167\\
 13.5&     27&     90&    173&    117&     29&      1&    437\\
 14.5&     39&    159&    447&    368&     96&      5&   1114\\
Total&     96&    321&    709&    537&    136&      7&   1807\\
\enddata
\end{deluxetable}

\begin{deluxetable}{crrrrrrr}
\tablecolumns{8}
\tabletypesize{\small}
\tablecaption{Expected Transiting CEGPs in \emph{Kepler's} Field of View
\label{ntranCEGPtbl}}
\tablewidth{0pt} 
\tablehead{ 
\colhead{}&\multicolumn{6}{c}{Spectral Type}&\colhead{}\\
\colhead{$\mr$} & \colhead{B5} & \colhead{A5}
& \colhead{F5} & \colhead{G5} & \colhead{K5} & \colhead{M5} & 
\colhead{All}}
\startdata 
  9.5&      0&      0&      0&      0&      0&      0&      1\\
 10.5&      0&      1&      1&      0&      0&      0&      2\\
 11.5&      1&      2&      2&      1&      0&      0&      6\\
 12.5&      2&      4&      6&      4&      1&      0&     17\\
 13.5&      3&      9&     17&     12&      3&      0&     44\\
 14.5&      4&     16&     45&     37&     10&      1&    111\\
Total&     10&     32&     71&     54&     14&      1&    181\\
\enddata
\end{deluxetable}

\begin{deluxetable}{cccccc}
\tablecolumns{6}
\tabletypesize{\small}
\tablecaption{Number of Expected Detections vs.  Assumed Number of
Fourier Components
\label{ndetfouriertbl}}
\tablewidth{0pt} 
\tablehead{ 
\colhead{}&\multicolumn{4}{c}{Atmospheric Model}&\colhead{}\\
\colhead{$M$} & \colhead{$\bar{r}=1.0$\tablenotemark{a}} &
\colhead{$\bar{r}=0.1$\tablenotemark{a}} & 
\colhead{$p=2/3$\tablenotemark{b}} & \colhead{$p=0.15$\tablenotemark{b}} &
\colhead{Average} } 
\startdata 
1  &  173.7  &  168.7  &  738.0  &  158.9  &  309.8  \\ 
2  &  184.7  &  155.3  &  736.6  &  146.9  &  305.9  \\ 
3  &  183.8  &  140.4  &  719.7  &  130.8  &  293.7  \\ 
4  &  175.0  &  126.7  &  706.6  &  117.6  &  281.5  \\ 
5  &  165.8  &  116.1  &  693.6  &  107.7  &  270.8  \\ 
6  &  159.1  &  108.6  &  683.2  &  101.0  &  263.0  \\ 
7  &  152.9  &  102.5  &  675.6  &   96.0  &  256.8  \\ 
\enddata \tablenotetext{a}{Atmospheric models from
\citet{seageretal2000} with mean particle radii $\bar{r}$ in microns.}
\tablenotetext{b}{Lambert sphere models with the given geometric
albedos, $p$.}
\end{deluxetable}

\begin{deluxetable}{crrrrcc}
\tabletypesize{\small}
\tablecolumns{7}
\tablecaption{Analytical Fits to Monte Carlo Null Distributions
\label{nfittbl}}
\tablewidth{0pt} \tablehead{ \colhead{}&\multicolumn{4}{c}{Fit to Single 
Test\tablenotemark{a}}&\colhead{Fit to}&\colhead{}\\
\colhead{}&\multicolumn{2}{c}{Direct Fit}&\multicolumn{2}{c}{Fit to Tail}&
\colhead{Single Search\tablenotemark{b}}&\colhead{}\\
\colhead{$M$}&\colhead{$A$}&\colhead{$B$}&\colhead{$A$}&
\colhead{$B$}&\colhead{$\neit$} &\colhead{Threshold\tablenotemark{c}}}

\startdata
1  &   2.110  &    2.114  &   1.923  &    2.691  &   451.81  &  16.9 \\ 
2  &   2.106  &    4.231  &   1.936  &    4.911  &   429.73  &  18.8 \\ 
3  &   2.104  &    6.346  &   2.001  &    6.738  &   462.57  &  20.0 \\ 
4  &   2.104  &    8.460  &   1.995  &    9.002  &   463.56  &  21.3 \\ 
5  &   2.103  &   10.574  &   2.006  &   11.082  &   469.40  &  22.3 \\ 
6  &   2.103  &   12.688  &   1.980  &   13.548  &   459.68  &  23.5 \\ 
7  &   2.104  &   14.801  &   2.037  &   15.170  &   476.03  &  24.1 \\ 
\enddata
\tablenotetext{a}{The fit is of the form 
$P_{\lone}(x,M)\approx P_{\chi^{2}_{2\,M}} \left(A\,x +B\right)$
}
\tablenotetext{b}{The fit is of the form
$P_{\lmax}(x,M)\approx P^{\neit}_{\chi^{2}_{2\,M}} \left(A\,x +B\right)$, 
where $A$ and $B$ are fits to the tail of the single test distributions.}
\tablenotetext{c}{Threshold for a false alarm rate of 1 in  
$10^{5}$ searches of stellar light curves.}
\end{deluxetable}

\begin{deluxetable}{crrrrrr}
\tablecolumns{7}
\tabletypesize{\small}
\tablecaption{Average Detection Rate for 1.2 R$_{\mathrm{J}}$ planets 
Orbiting Sun-Like Stars,
(\%)
\label{avgDRtbl} }
\tablewidth{0pt} 
\tablehead{ \colhead{}&
\multicolumn{6}{c}{Apparent Stellar Magnitude}\\
\colhead{P$_{rot}$}&\multicolumn{6}{c}{($\mr$)}\\
\colhead{(Days)} & \colhead{9.5} & \colhead{10.5} & \colhead{11.5}
& \colhead{12.5} & \colhead{13.5} & \colhead{14.5} } 
\startdata 
\cutinhead{\small{$\bar r$=1.0 $\mu$m Particles}}
 5&    0.0&    0.0&    0.0&    0.0&    0.0&    0.0\\
10&    0.0&    0.0&    0.0&    0.0&    0.0&    0.0\\
15&    0.0&    0.0&    0.0&    0.0&    0.0&    0.0\\
20&   12.2&   12.0&   11.8&   10.8&    8.2&    2.6\\
25&   36.0&   35.7&   34.6&   31.8&   24.0&    8.2\\
30&   49.6&   48.7&   47.4&   43.5&   33.2&   13.3\\
35&   59.3&   58.2&   55.3&   53.0&   40.8&   15.9\\
40&   66.5&   65.9&   64.4&   56.6&   44.6&   16.8\\
\cutinhead{\small{$\bar r$=0.1 $\mu$m Particles}}
 5&    0.0&    0.0&    0.0&    0.0&    0.0&    0.0\\
10&    0.0&    0.0&    0.0&    0.0&    0.0&    0.0\\
15&    0.0&    0.0&    0.0&    0.0&    0.0&    0.0\\
20&   10.8&   10.6&   10.3&    9.9&    5.1&    0.0\\
25&   36.5&   36.3&   35.7&   34.0&   25.8&    5.0\\
30&   53.5&   53.2&   51.6&   48.3&   39.2&    9.5\\
35&   62.9&   62.1&   60.4&   58.2&   46.9&   10.0\\
40&   72.0&   71.5&   68.8&   64.4&   51.1&   10.2\\
\cutinhead{\small{Albedo $p=0.15$ Lambert Sphere}}
 5&    0.0&    0.0&    0.0&    0.0&    0.0&    0.0\\
10&    0.0&    0.0&    0.0&    0.0&    0.0&    0.0\\
15&    0.0&    0.0&    0.0&    0.0&    0.0&    0.0\\
20&    6.8&    6.7&    6.3&    4.9&    1.0&    0.0\\
25&   38.6&   38.4&   37.5&   34.0&   25.4&    1.2\\
30&   56.6&   56.4&   55.9&   52.7&   42.4&    4.6\\
35&   67.3&   67.1&   65.6&   61.2&   50.0&    5.6\\
40&   75.6&   75.4&   74.4&   70.1&   54.7&    5.9\\
\cutinhead{\small{Albedo $p=2/3$ Lambert Sphere}}
 5&    0.0&    0.0&    0.0&    0.0&    0.0&    0.0\\
10&    0.0&    0.0&    0.0&    0.0&    0.0&    0.0\\
15&   39.0&   39.0&   39.0&   38.9&   38.8&   37.5\\
20&   67.1&   67.1&   67.0&   66.9&   66.3&   64.3\\
25&   82.4&   82.4&   82.4&   82.3&   81.9&   78.8\\
30&   84.1&   84.1&   84.1&   84.1&   83.6&   80.9\\
35&   93.9&   93.9&   93.8&   93.4&   92.4&   84.6\\
40&   97.3&   97.3&   97.2&   96.4&   95.6&   89.1\\
\enddata
\end{deluxetable}

\begin{deluxetable}{rrrrrrrr}
\tabletypesize{\small}
\tablecolumns{8}
\tablecaption{Number of CEGPs Detected, Signal Adjusted for Stellar 
Radius
\label{nrefltbla}}
\tablewidth{0pt} 
\tablehead{ \colhead{}                        &
            \multicolumn{6}{c}{Spectral Type} & 
            \colhead{}                        \\
            \colhead{$\mr$}                   & 
			\colhead{B5}                      & 
			\colhead{A5}                      & 
			\colhead{F5}                      &
			\colhead{G5}                      & 
			\colhead{K5}                      & 
			\colhead{M5}                      & 
			\colhead{All}                     }
\startdata 
\cutinhead{\scriptsize{$\bar r$=1.0 $\mu$m Particles}}
  9.5&      0&      0&      0&      0&      0&      0&      1\\
 10.5&      0&      0&      0&      1&      1&      0&      2\\
 11.5&      0&      0&      0&      5&      2&      0&      7\\
 12.5&      0&      0&      1&     14&      7&      0&     23\\
 13.5&      0&      0&      2&     37&     22&      1&     62\\
 14.5&      0&      0&      0&     54&     60&      5&    120\\
Total&      0&      0&      3&    112&     91&      7&    214\\
\cutinhead{\scriptsize{$\bar r$=0.1 $\mu$m Particles}}
  9.5&      0&      0&      0&      0&      0&      0&      1\\
 10.5&      0&      0&      0&      1&      1&      0&      2\\
 11.5&      0&      0&      0&      5&      2&      0&      7\\
 12.5&      0&      0&      1&     16&      7&      0&     24\\
 13.5&      0&      0&      2&     42&     23&      1&     68\\
 14.5&      0&      0&      0&     50&     64&      5&    119\\
Total&      0&      0&      3&    115&     96&      7&    222\\
\cutinhead{\scriptsize{Lambert Sphere, $p$=0.15}}
  9.5&      0&      0&      0&      0&      0&      0&      1\\
 10.5&      0&      0&      0&      2&      1&      0&      2\\
 11.5&      0&      0&      0&      5&      2&      0&      8\\
 12.5&      0&      0&      1&     17&      7&      0&     25\\
 13.5&      0&      0&      1&     46&     22&      1&     71\\
 14.5&      0&      0&      0&     45&     60&      5&    110\\
Total&      0&      0&      3&    115&     92&      7&    216\\
\cutinhead{\scriptsize{Lambert Sphere, $p$=2/3}}
  9.5&      0&      0&      0&      1&      0&      0&      1\\
 10.5&      0&      0&      1&      3&      1&      0&      4\\
 11.5&      0&      0&      4&      9&      2&      0&     15\\
 12.5&      0&      0&     13&     29&      8&      0&     50\\
 13.5&      0&      0&     38&     93&     27&      1&    160\\
 14.5&      0&      0&    104&    283&     90&      5&    482\\
Total&      0&      0&    160&    417&    128&      7&    712\\

\enddata
\end{deluxetable}

\begin{deluxetable}{rrrrrrrr}
\tabletypesize{\small}
\tablecolumns{8}
\tablecaption{Number of CEGPs Detected, Signal Not Adjusted for 
Stellar Radius
\label{nrefltblb}}
\tablewidth{0pt} 
\tablehead{ \colhead{}                        &
	    \multicolumn{6}{c}{Spectral Type} & 
	    \colhead{}                        \\
	    \colhead{$\mr$}                   & 
			\colhead{B5}                      & 
			\colhead{A5}                      & 
			\colhead{F5}                      &
			\colhead{G5}                      & 
			\colhead{K5}                      & 
			\colhead{M5}                      & 
			\colhead{All}                     }
\startdata 

\cutinhead{\scriptsize{$\bar{r}$=1.0 $\mu$m Particles}}
  9.5&      0&      0&      0&      0&      0&      0&      0\\
 10.5&      0&      0&      0&      1&      0&      0&      2\\
 11.5&      0&      0&      1&      4&      1&      0&      6\\
 12.5&      0&      0&      2&     13&      3&      0&     19\\
 13.5&      0&      0&      6&     32&      9&      1&     48\\
 14.5&      0&      0&      6&     40&     11&      1&     59\\
Total&      0&      0&     16&     91&     25&      2&    133\\
\cutinhead{\scriptsize{$\bar{r}$=0.1 $\mu$m Particles}}
  9.5&      0&      0&      0&      0&      0&      0&      0\\
 10.5&      0&      0&      0&      1&      0&      0&      2\\
 11.5&      0&      0&      1&      5&      1&      0&      6\\
 12.5&      0&      0&      3&     14&      4&      0&     21\\
 13.5&      0&      0&      5&     36&     10&      1&     52\\
 14.5&      0&      0&      3&     24&      7&      1&     34\\
Total&      0&      0&     12&     81&     22&      2&    116\\
\cutinhead{\scriptsize{Lambert Sphere, $p$=0.15}}
  9.5&      0&      0&      0&      0&      0&      0&      0\\
 10.5&      0&      0&      0&      1&      0&      0&      2\\
 11.5&      0&      0&      1&      5&      1&      0&      7\\
 12.5&      0&      0&      2&     15&      4&      0&     21\\
 13.5&      0&      0&      4&     38&     10&      1&     53\\
 14.5&      0&      0&      1&     13&      3&      0&     18\\
Total&      0&      0&      8&     72&     19&      1&    102\\
\cutinhead{\scriptsize{Lambert Sphere, $p$=2/3}}
  9.5&      0&      0&      1&      1&      0&      0&      1\\
 10.5&      0&      0&      2&      2&      1&      0&      5\\
 11.5&      0&      0&      6&      9&      2&      0&     16\\
 12.5&      0&      0&     19&     28&      7&      0&     55\\
 13.5&      0&      0&     58&     89&     24&      1&    173\\
 14.5&      0&      0&    161&    272&     76&      5&    514\\
Total&      0&      0&    247&    400&    110&      6&    764\\

\enddata
\end{deluxetable}

\begin{deluxetable}{lrrrrrr}
\tabletypesize{\scriptsize}
\tablecaption{Number of Background Binaries Not Excluded by Astrometry
\label{nbackbintbl}}
\tablewidth{0pt} 
\tablehead{ \colhead{ } & \multicolumn{6}{c}{Apparent
Stellar Magnitude}\\
\colhead{ } & \multicolumn{6}{c}{($\mr$)}\\
\colhead{Parameter}& \colhead{9.5} & \colhead{10.5} & \colhead{11.5} &
\colhead{12.5} & \colhead{13.5} & \colhead{14.5} }

\startdata 
Number of Background Binaries in Target Apertures\tablenotemark{a}&
3& 18& 85& 296& 903& 2405\\
Centroid Rejection Radius (arcsec)\tablenotemark{b}& 
$<$0.5&$<$0.5& $<$0.5& $<$0.5& $<$0.5& 0.7\\
Aperture Size (square arcsec)&  400&  384&  352&  288&  240&  192\\
Number of Potential False Alarms\tablenotemark{c}&  0& 0& 0& 1& 3& 18 \\

\enddata \tablenotetext{a}{\scriptsize The background binaries of
concern have periods between 2 and 7 days and are 8 magnitudes fainter
than the target stars.  See Table \ref{nstartbl} for the number of
target stars in each magnitude bin.} \tablenotetext{b}{\scriptsize
Background variables can be rejected outside this radius with a
confidence level of 99.9\%.} \tablenotetext{c}{\scriptsize These are
the expected numbers of background variables that cannot be rejected
simply by examining \emph{Kepler} data.  Follow-up observations may be
necessary to distinguish them from CEGPs if the objects display
coherent, periodic light curves over the 4 yr duration of
\emph{Kepler's} observations.}
\end{deluxetable}

\end{document}